\documentclass[%
 reprint,
nofootinbib,
 amsmath,amssymb,
 aps,
]{revtex4-2}
\usepackage{graphicx}
\usepackage{dcolumn}
\usepackage{bm}
\usepackage{hyperref}

\usepackage{orcidlink}

\usepackage{physics}

\usepackage[normalem]{ulem} 

\usepackage{aas_macros}

\usepackage{subfigure}

\usepackage{booktabs} 
\usepackage{siunitx}  
\usepackage{caption} 

\begin{document}

\preprint{APS/123-QED}

\title{Collective Thomson scattering in magnetized electron and positron pair plasma \\and the application to induced Compton scattering}

\author{Rei Nishiura\orcidlink{0009-0003-8209-5030}}
 \email{nishiura@tap.scphys.kyoto-u.ac.jp}
 \affiliation{%
 Department of Physics, Kyoto University, Kyoto 606-8502, Japan}%
\author{Kunihito Ioka\orcidlink{0000-0002-3517-1956}}%
 \email{kunihito.ioka@yukawa.kyoto-u.ac.jp}
\affiliation{%
 Center for Gravitational Physics and Quantum Information, 
 Yukawa Institute for Theoretical Physics, Kyoto University, Kyoto 606-8502, Japan}%

%
%

\date{\today}

\begin{abstract}
We consider collective Thomson scattering (CTS) of an incident X-mode wave (with the electric vector perpendicular to the background magnetic field) in magnetized electron and positron pair plasma. The collective effects do not exactly cancel out in contrast to the non-magnetized case. Still, the cross-section is comparable to the non-collective one, with the same suppression by the square of the cyclotron frequency in a strong magnetic field. The comparable cross-section holds even though the net current is nearly zero from the drift motion of electrons and positrons. The plasma response does not also affect the cross-section so much. 
The spectrum of the scattered wave in finite temperature plasma peaks at cyclotron overtones.
Based on these results, we also estimate induced Compton scattering in strongly magnetized pair plasma. Implications for pulsars and fast radio bursts are discussed.
\end{abstract}

\maketitle


\section{INTRODUCTION}
Fast Radio Bursts (FRBs) are the brightest radio transients, first discovered in 2007~\citep{2007Sci...318..777L,2013Sci...341...53T,2019A&ARv..27....4P}; most FRBs originate outside our Galaxy, and their origin is not fully understood. Notably, the observation of FRB 20200428 in 2020, coinciding with an X-ray burst from the Galactic magnetar SGR 1935+2154~\citep{2020Natur.587...54C,2020Natur.587...59B,2020ATel13687....1Z,2020ApJ...898L..29M,2020ATel13686....1T,2020ATel13688....1R}, marked a step towards understanding these phenomena. FRBs are also used as a tool to extract information from the traces of interactions of FRBs with distant intergalactic material during their propagation and to apply this information to cosmology \citep{2003ApJ...598L..79I,2004MNRAS.348..999I,2020Natur.581..391M} (see the review by \citet{2021Univ....7...85B} for various studies).

Despite advancements in observational and applicational studies, the emission mechanism of FRBs remains unresolved. The emission region of FRBs has sparked a debate \citep{2020MNRAS.498.1397L,2020arXiv200505093L,2020ApJ...900L..21Y,2021MNRAS.500.2704Y} regarding whether they arise within a magnetosphere of the magnetar \citep{2014PhRvD..89j3009K,2016MNRAS.457..232C,2016MNRAS.462..941L,2017MNRAS.468.2726K,2017ApJ...836L..32Z,2018MNRAS.477.2470L,2018A&A...613A..61G,2018ApJ...868...31Y,2020ApJ...893L..26I,2020MNRAS.494.2385K,2021MNRAS.508L..32C} or through interactions between circumstellar matter located at a distance from the magnetosphere and relativistic outflows from the magnetar \citep{2014MNRAS.442L...9L,2016MNRAS.461.1498M,2017ApJ...842...34W,2020MNRAS.494.4627M,2019MNRAS.485.4091M,2020ApJ...896..142B,2021MNRAS.500.2704Y}. It has been argued that coherent waves such as FRBs and radio emission from pulsars may cause induced Compton scattering with electron and positron plasma~($e^{\pm}$ plasma) in the magnetosphere and may not escape the magnetosphere if the Lorentz factor of the scattered particles is small ~\citep{1975Ap&SS..36..303B,1978MNRAS.185..297W,1982MNRAS.200..881W,2008ApJ...682.1443L}.

Induced Compton scattering is a process in which the reaction rate increases due to the induced effect, primarily when there is a high occupation number of incident photons (coherent light)~\citep{1975Ap&SS..36..303B,1978MNRAS.185..297W,1982MNRAS.200..881W,2008ApJ...682.1443L}. In particular, even with low photon occupancy in background radiation, the scattering can exponentially amplify the background radiation and significantly attenuate the incident light if the incident wave is highly collimated \citep{1996AstL...22..399L}. Theoretically, it is described by incorporating the Compton scattering collision term into the Boltzmann equation for scattered photons. As the first-order quantum correction cancels out, this process is understood classically \citep{1974PhFl...17..778D,1974PhFl...17.1757O,1975PhFl...18..201L,1982MNRAS.200..881W,2022ApJ...930..106G} \footnote{The classical interpretation of induced Compton scattering is often called "induced Thomson scattering" \citep{1972A&A....19..135L,1977PhFl...20.1674J,1978PhFl...21..404C,1979PhFl...22.1115C}. In astrophysics, it is common for both classical and quantum mechanical interpretations to be encompassed under the term "induced Compton scattering"\citep{1974PhFl...17..778D,1974PhFl...17.1757O,1975PhFl...18..201L,1982MNRAS.200..881W,2022ApJ...930..106G}.}. The classical interpretation is that plasma density fluctuations are produced by the beat between incident and scattered electromagnetic waves.

This paper focuses on three effects that can influence the scattering processes to understand the observed FRBs. The first effect is the suppression of Thomson scattering due to a strong magnetic field. In the presence of a strong magnetic field, the motion of scattering particles is constrained by the magnetic field, leading to a reduced Thomson scattering cross-section for X-mode electromagnetic waves~\citep{1971PhRvD...3.2303C,1979PhRvD..19.2868H,1979PhRvD..19.1684V,1992herm.book.....M,2000ApJ...540..907G} \footnote{In $e^{\pm}$ plasma, an X-mode wave is a linearly polarized electromagnetic wave whose electric field component is perpendicular to the plane formed by the background magnetic field and the wave's propagation direction. The other O-mode wave has its electric field component within this plane.}. Consequently, the rate for induced Compton scattering may also be suppressed \citep{2020MNRAS.494.1217K}. The second effect is the plasma response to radiation. \citet{2004ApJ...600..872G} estimated the curvature radiation from charged particles moving along an infinitely strong curved magnetic field, considering the plasma response. They found that the radiation is significantly suppressed compared to vacuum curvature radiation. This fact implies the need to consider plasma response also in Thomson scattering. Regarding the third effect, there is an intuitive argument that the electric field of the incident electromagnetic wave causes electrons and positrons to drift in the same direction, leading to a mutual cancellation of currents and significant suppression of scattering in a strong magnetic field \citep{2020ApJ...897....1L}. 

It is crucial to consider these effects consistently to treat the scattering processes in the magnetar magnetosphere. These effects alter the reaction rate of induced Compton scattering in the scattering cross-section. Therefore, we aim to cohesively integrate these effects into the Thomson cross-section for magnetized \(e^{\pm}\) plasma.

We believe that CTS can provide a unified explanation for both magnetic field effects and plasma effects (see the book \citet{2012FuST...61..104F} for CTS). 
CTS describes the generation of scattered waves through the interaction of electromagnetic waves propagating in plasma with plasma waves or density fluctuations of ions and electrons. This theory considers interactions among numerous charged particles (the analog of Debye screening in a thermal plasma) and is used for exploring physical properties of plasma through scattering induced by the screening field around ions \citep{1960CaJPh..38.1114F,1960RSPSA.259...79D,1960PhRv..120.1528S}.
CTS has been widely studied in the field of plasma physics \citep{hutchinson_2002,2012FuST...61..104F} and applied to precise measurements of ion temperature in laboratory plasma experiments \citep{1983IJIMW...4..205W,PhysRevLett.62.2833,2019HEDP...32...82B,2020PhPl...27j3104S}. However, most of the existing research in this theory deals with scattering in ion-electron plasma. Thomson scattering in $e^{\pm}$ plasma has been studied without a background magnetic field \citep{1992PhFlB...4.2669S}. \citet{1992PhFlB...4.2669S} showed that electrons and positrons completely cancel the collective effect. However, the collective scattering behavior in the presence of a background magnetic field has yet to be studied as far as we know.

CTS is a distinct theoretical concept from induced Compton scattering. In CTS theory, scattered electromagnetic waves are generated by the beat between the incident electromagnetic wave and pre-existing density fluctuations. In contrast, for induced Compton scattering,  density fluctuations arise from the beat between incident and scattered electromagnetic waves. In this paper, we employ CTS to calculate the Compton scattering cross-section per particle in the low-frequency limit and apply it to induced Compton scattering by assuming that the CTS picuture is also valid for the induced process.
 
This study is the first to investigate CTS in magnetized \( e^{\pm} \) plasma, focusing specifically on the scattering of X-mode waves. Magnetar magnetospheres are believed to contain \( e^{\pm} \) plasma and strong magnetic fields. Even in a strong magnetic field, electromagnetic waves with electric field components parallel to the magnetic field scatter electrons with the ordinal Thomson cross-section. The parallel electric field also accelerates \( e^{\pm} \), leading to Landau damping of the wave's energy. However, waves with electric field components perpendicular to the magnetic field have significantly suppressed cross-section for scattering, allowing more efficient propagation (see section \ref{sec:Thomson_free}). In addition, transverse propagation across the magnetic field is necessary for electromagnetic waves to escape the magnetosphere. Radio frequencies are usually lower in the magnetosphere than plasma and cyclotron frequencies. O-mode waves with low frequencies correspond to Alfvén waves, which carry energy along a magnetic field and are confined within closed magnetic fields. On the other hand, X-mode waves can propagate across the magnetic field while maintaining a perpendicular electric field component, which is often considered in the context of FRB and pulsar emissions \citep{1986ApJ...302..120A,2017RvMPP...1....5M,2017MNRAS.468.2726K,2020arXiv200109210L,2020MNRAS.494.1217K}. 

In Section \ref{sec:Thomson_free}, we review single-particle scattering, i.e., scattering of a free particle, in the presence of a background magnetic field. In Section \ref{sec:Thomson_scattering_from_a_magnetized_plasma}, we consider the Thomson scattering in magnetized $e^{\pm}$ plasma and the properties of the obtained scattering cross-section. Section \ref{sec:discussion_collective} discusses whether electrons and positrons cancel the Thomson scattering in a strong magnetic field. We also discuss a possible implication for observations of pulsars, taking the obtained spectra of the scattering cross-section into account. Furthermore, based on the analysis of the CTS, we estimate the effective optical depth of induced Compton scattering in a strong magnetic field. The discussion of Thomson scattering considering the plasma response is included in Appendix \ref{Ap:plasma_reaction} because it does not affect the main conclusions of this study. 

Throughout this paper, the notation \( A = 10^n A_n \) and the Centimeter-Gram-Second (CGS) system of units are consistently employed.

\section{THOMSON SCATTERING by a FREE PARTICLE IN A STRONG MAGNETIC FIELD}
\label{sec:Thomson_free}
We calculate Thomson cross-section for X-mode waves (linearly polarized perpendicular to the plane of the magnetic field and wave vector) in the presence of a strong magnetic field. We impose the following assumptions in deriving the details.
\begin{itemize} 
\item Consider a free electron or a free positron as a scattering particle and assume that it is static at the origin before scattering.
\item A uniform magnetic field $\bm{B}_0=(B_0,0,0)$ exists in the $x$-axis direction.
\item The X-mode wave is incident on the magnetic field with a wave-number vector \(\bm{k}_0\) and an angular frequency \(\omega_0\).
\item The magnetic field of the incident electromagnetic wave is assumed to be sufficiently small compared to the background magnetic field, that is $|\bm{B}_{\text{wave}}|\ll|\bm{B}_0|.$
\item The motion of a particle in the wave field is approximated as non-relativistic.
\end{itemize}

We neglect relativistic effects as an initial step because even a non-relativistic CTS framework has yet to be discussed for \( e^{\pm} \) plasma. Observations also suggest non-relativistic plasma environments in magnetar magnetospheres, exemplified by the observation of FRB 20200428 concurrent with a magnetar X-ray burst with non-relativistic energy of $\sim$80 keV \citep{2020ApJ...898L..29M}. Theoretical studies further indicate that an \( e^{\pm} \) fireball generated by the X-ray burst has a sub-relativistic plasma temperature in the comoving frame, despite the bulk of the fireball being accelerated to relativistic speeds \citep{2020ApJ...904L..15I,2023MNRAS.519.4094W}. The plasma within the fireball is also thermalized in the comoving frame. Therefore, our theory can be applied at least in the comoving frame of the fireball plasma.

The electric field of the X-mode plane wave in pair plasma can be written as
\begin{equation}
E_{\mathrm{X}}^{\mathrm{in}}(t)=E_0 e^{-i \omega_0 t}\left(\begin{array}{l}
0 \\
1 \\
0
\end{array}\right).
\end{equation}
Then the equation of motion for particles in the wave field can be denoted by
\begin{equation}
m_{\mathrm{e}} \dot{\boldsymbol{v}}_{ \pm}= \pm e \boldsymbol{E}_{\mathrm{X}}^{\mathrm{in}} \pm \frac{e}{c} \boldsymbol{v}_{ \pm} \times \boldsymbol{B}_0,
\end{equation}
where $e$ is the absolute value of elementary charge (i.e., the positron charge) and $c$ is the speed of light. From this equation of motion, the motion of the particle is represented by 
\begin{equation}
\boldsymbol{v}_{ \pm}=\frac{e E_0}{m_{\mathrm{e}} \omega_0} \frac{\omega_0^2}{\omega_0^2-\omega_{\mathrm{c}}^2}\left(\begin{array}{c}
0 \\
\pm i \\
\frac{\omega_{\mathrm{c}}}{\omega_0}
\end{array}\right) e^{-i \omega_0 t},
\label{eq:velocity_of_moving_particles}
\end{equation}
where
\begin{equation}
    \omega_{\mathrm{c}}\equiv\frac{eB_0}{m_{\text{e}}c}
\end{equation}
is the electron cyclotron frequency. A strong magnetic field means that the cyclotron frequency is sufficiently large compared to the angular frequency of the incident electromagnetic wave~($\omega_{\text{c}}\gg\omega_0$)~.

When the background magnetic field is strong, the particle motion is characterized by a dominant drift motion. The particles in the wave field have a figure-8 motion in the plane perpendicular to the background magnetic field. When the background field is strong, the drift velocity is $(\omega_{\text{c}}/\omega_0)$ times larger than that in the direction of the incident electric field. The physical reason for this is that the particles, supposed initially to oscillate in the direction of the incident electric field, are immediately bent to the drift direction by the strong background magnetic field.

The electromagnetic field produced by the oscillating particles is described by Liénard–Wiechert potentials
\begin{equation}
\begin{aligned}
&\boldsymbol{E}_{\mathrm{rad}} \equiv \frac{q}{c R}\left[ \boldsymbol{n} \times(\boldsymbol{n} \times \dot{\boldsymbol{\beta}})\right]_{\text{ret}},   \\
&\boldsymbol{B}_{\mathrm{rad}} \equiv \frac{q}{c R}\left[ \boldsymbol{n} \times\left\{\bm{n}\times(\boldsymbol{n} \times \dot{\boldsymbol{\beta}})\right\}\right]_{\text{ret}},
\end{aligned}
\label{eq:Riena-ru}
\end{equation}
where $\bm{\beta}\equiv\bm{v}/c$ and the retarded time is defined as follows 
\begin{equation}
t^{\prime}=t-\frac{|\bm{R}-\bm{r}\left(t^{\prime}\right)|}{c}.
\end{equation}
Let $\bm{R}$ be the observer's position and $\bm{n}$ be the unit vector from the charged particle to the observer at the retarded time. If the observer is sufficiently far away from the radiation source, the retarded time can be approximated as
\begin{equation}
t^{\prime} \simeq t-\frac{R}{c}+\frac{\bm{n} \cdot \bm{r}}{c}.
\end{equation}

The energy radiated by the oscillating particles per unit time can be calculated by the radiative Poynting flux through a sphere of sufficiently large radius. In the non-relativistic limit, this is expressed as
\begin{equation}
P_{\mathrm{NR}}=\frac{e^2}{4 \pi c} \int \mathrm{d} \Omega|\boldsymbol{n} \times(\boldsymbol{n} \times \dot{\boldsymbol{\beta}})|^2.
\label{eq:Radiative_power_NR}
\end{equation}
If the angle between $\bm{n}$ and $\dot{\bm{\beta}}$ is $\theta$, it can be written as $|\boldsymbol{n} \times(\boldsymbol{n} \times \dot{\boldsymbol{\beta}})|^2=\dot{\beta}^2 \sin ^2 \theta$.

Note that when the plasma density is enormous (i.e., $\omega_{\text{p}}\gg\omega_0$), the response of the plasma must be taken into account, and the Liener-Wiechert potentials, which assumes electromagnetic wave propagation in a vacuum, cannot be used. In Appendix \ref{Ap:plasma_reaction}, we estimated the energy of scattered X-mode electromagnetic waves by particles in the limit of large plasma density and background magnetic field (i.e., the limit of $\omega_{\text{c}},\omega_{\text{p}}\gg\omega_0$). However, even considering the plasma response, the scattering cross-section of the X-mode wave is found to be within $50\%$ of that in vacuum for
\begin{equation}
\omega_{\text{c}}>\omega_{\text{p}}\gg\omega_0.     
\end{equation}
Therefore, we adopt Liener-Wiechert potentials to evaluate the order of the scattering cross-section in this study. Computing radiation using plasma wave modes is a future work.

Substituting the motion of the oscillating particle into equation \eqref{eq:Radiative_power_NR}, we obtain the energy per unit time emitted from an electron or a positron in the X-mode waves
\begin{equation}
\left\langle\frac{\mathrm{d} P_{\mathrm{X}}}{\mathrm{d} \Omega}\right\rangle=\frac{e^4 E_0^2}{8 \pi m_{\mathrm{e}}^2 c^3}\left(\frac{\omega_0^2}{\omega_0^2-\omega_{\text{c}}^2}\right)^2\left\{1+\left(\frac{\omega_{\text{c}}}{\omega_0}\right)^2\right\} \sin ^2 \theta.
\end{equation}
The scattering cross-section of the X-mode waves is obtained by dividing the scattered energy per unit time by the energy flux of the incident electromagnetic wave
\begin{equation}
\sigma_{\mathrm{X}}=\frac{8 \pi}{c E_0^2}\left\langle P_{\mathrm{X}}\right\rangle=\frac{1}{2} \sigma_{\mathrm{T}}\left\{\left(\frac{\omega_0}{\omega_0+\omega_{\mathrm{c}}}\right)^2+\left(\frac{\omega_0}{\omega_0-\omega_{\mathrm{c}}}\right)^2\right\},
\label{eq:one_particle_cross_section}
\end{equation}
where
\begin{equation}
    \sigma_{\mathrm{T}}\equiv\frac{8\pi}{3}r_{\text{e}}^2\equiv\frac{8\pi}{3}\left(\frac{e^2}{m_{\text{e}}c^2}\right)^2
\end{equation}
is Thomson cross-section.

If the background magnetic field is sufficiently large, the scattering of an X-mode wave is suppressed by a factor of $(\omega_0/\omega_{\text{c}})^2$. The physical interpretation is that 
although the electric field of the X-mode wave tries to swing the charged particle, the charged particle firmly sticks to the background magnetic field and is hardly shaken by the waves. As a result, the radiation from the particle is suppressed.

\section{THOMSON SCATTERING IN ELECTRON-POSITRON MAGNETIZED PLASMA}
\label{sec:Thomson_scattering_from_a_magnetized_plasma}
This section considers Thomson scattering of electromagnetic waves by $e^{\pm}$ plasma. The following are the differences in the setup from the previous section.
\begin{itemize}
\item Assume $e^{\pm}$ plasma as a scattering medium.
\item For simplicity, we assume that only longitudinal wave components are produced by density fluctuations in the $e^{\pm}$ plasma~(electrostatic approximation). It has been argued that without accounting for the full electromagnetic fluctuations, resonance peaks due to electromagnetic waves would not be visible in the scattering spectra \citep{1992NucFu..32..745A,2012FuST...61..104F}.
\end{itemize}
\subsection{Basic equations}
This section formally derives the energy radiated per unit time by $e^{\pm}$ plasma when it scatters electromagnetic waves. First, the equation of motion of an electron or a positron perturbed by an X-mode electromagnetic wave is given by
\begin{equation}
m_{\text{e}} \boldsymbol{\dot{v}}_{ \pm}= \pm e \boldsymbol{E}_{i0}e^{i\left(\bm{k}_{0}\cdot\bm{r}\left(t^{\prime}\right)-\omega_0t^{\prime}\right)} \pm \frac{e}{c} \boldsymbol{v}_{ \pm} \times \boldsymbol{B}_0.
\label{eq:EOM_for_plasma}
\end{equation}
Solving this equation yields the velocity of the oscillating particles as follows
\begin{equation}
\boldsymbol{v}_{ \pm}=\frac{e E_0}{m_{\text{e}} \omega_0} \frac{\omega_0^2}{\omega_0^2-\omega_{\mathrm{c}}^2}\left(\begin{array}{c}
0 \\
\pm i \\
\frac{\omega_{\text{c}}}{\omega_0}
\end{array}\right) e^{i\left(\bm{k}_{0}\cdot\bm{r}-\omega_0t\right)}.
\label{eq:motion_of_pm}
\end{equation}

In considering plasma scattering, it has been argued that scattering from the uniform density component can be neglected \citep{bekefi1966radiation}. As a simple physical interpretation, for a scattered wave emitted in a specific direction and wavelength, imagine a pair of thin uniform plasma plates, separated by half the wavelength, aligned perpendicular to the direction of wave travel. Since the phases of the scattered waves from the thin plates are out of phase by $\pi$, the scattered waves cancel each other perfectly. Considering similar pairs over the entire scattering region, all scattered waves from the uniform density component can be neglected. In other words, all the scattering of electromagnetic waves in the plasma is caused by statistical density fluctuations.

The scattered electric field by the electron or positron population can be evaluated by Liénard–Wiechert potentials produced by the density fluctuations
\begin{equation}
\begin{aligned}
\bm{E}_{ \pm}(\bm{R}, t)&= \pm \frac{e}{c R} \int_V \dd^3 \bm{r} \int \dd^3 \bm{v}~ \delta F_{ \pm}\left(\bm{r}, \bm{v}, t^{\prime}\right)\\
&\times\left[ \boldsymbol{n} \times(\boldsymbol{n} \times \dot{\boldsymbol{\beta}}_{\pm})\right]_{\text{ret}} .   
\end{aligned}
\end{equation}
The relationship between the first order perturbation of the distribution function in the scattering region, $\delta F_{ \pm}$, and the density fluctuations $\delta n_{\pm}(\bm{r},t)$ is
\begin{equation}
    \delta n_{\pm}(\bm{r},t)=\int\dd^3\bm{v}~\delta F_{ \pm}(\bm{r},\bm{v},t).
    \label{eq:density_fluctuation15}
\end{equation}
The total electric field scattered by the plasma is the sum of the contributions from the electron and positron populations
\begin{equation}
\begin{aligned}
\bm{E}_{ \text{tot}}(\bm{R}, t)&=\frac{e}{c R} \int_V \dd^3 \bm{r} \int \dd^3 \bm{v}~\\
&\times\left\{\delta F_{ +}\left(\bm{r}, \bm{v}, t^{\prime}\right)\left[ \boldsymbol{n} \times(\boldsymbol{n} \times \dot{\boldsymbol{\beta}}_{+})\right]_{\text{ret}}\right.\\
& \left.-\delta F_{-}\left(\bm{r}, \bm{v}, t^{\prime}\right)\left[ \boldsymbol{n} \times(\boldsymbol{n} \times \dot{\boldsymbol{\beta}}_{-})\right]_{\text{ret}}\right\}.
\end{aligned}
\label{eq:total_electric_field}
\end{equation}

The energy radiated per unit time and unit solid angle can be time averaged over a large enough sphere
\begin{equation}
\frac{\dd P_{\text{s}}}{\dd \Omega}(\bm{R}) =\frac{c R^2}{4 \pi} \lim _{T \rightarrow \infty} \frac{1}{T} \int_{-\frac{T}{2}}^{\frac{T}{2}} \dd t~\left|\bm{E}_{\text{tot} }(\bm{R}, t)\right|^2.
\end{equation}
Here the time component of the scattered electric field is Fourier-transformed into its frequency component
\begin{equation}
\widetilde{\bm{E}_{\text{tot}}}\left(\bm{R},\omega_1\right)=\int_{-\infty}^{+\infty} \dd t~ \bm{E}_{\text{tot} }(\bm{R},t) e^{-i \omega_1 t}.
\label{eq:Fourier_transformed_electric_field}
\end{equation}
Using Parseval's identity for the absolute square of the radiative electric field, the radiation power per solid angle can be expressed as
\begin{equation}
\frac{\dd P_{\text{s}}}{\dd \Omega}(\bm{R})=\frac{c R^2}{4 \pi}  \lim _{T \rightarrow \infty} \frac{1}{\pi T} \int_{0}^{\infty} \dd \omega_1~\left|\widetilde{\bm{E}_{\text{tot}}}\left(\omega_1\right)\right|^2.
\label{eq:radiation_power_per_solid_angle}
\end{equation}

In the subsequent section, we will calculate the scattered electric field from $e^{\pm}$ plasma in an X-mode wave specifically.
\subsection{Radiative electric field}
In this section, we calculate the electric field radiated from $e^{\pm}$ plasma in an X-mode wave and show that it can be written as a combination of density fluctuations. The density fluctuation is evaluated in the next section. From equations \eqref{eq:density_fluctuation15}, \eqref{eq:total_electric_field}, and \eqref{eq:Fourier_transformed_electric_field}, the Fourier-transformed scattered electric field is obtained by adding up the electric fields created by the electron and positron density fluctuations at the retarded time over the scattering region as follows
\begin{equation}
\begin{aligned}
\widetilde{\bm{E}_{\text{tot}}}\left(\omega_1\right)&=\frac{e}{c R}\int_{-\infty}^{\infty}\dd t \int_V \dd^3 \bm{r}~ e^{-i\omega_1\left(t^{\prime}+\frac{R}{c}-\frac{\bm{n} \cdot \bm{r}}{c}\right)}~\\
&\times\left\{\delta n_{ +}\left(\bm{r},t\right)\left[ \boldsymbol{n} \times(\boldsymbol{n} \times \dot{\boldsymbol{\beta}}_{+})\right]_{\text{ret}}\right.\\
& \left.-\delta n_{-}\left(\bm{r},t\right)\left[ \boldsymbol{n} \times(\boldsymbol{n} \times \dot{\boldsymbol{\beta}}_{-})\right]_{\text{ret}}\right\},
\end{aligned}
\label{eq:scattered_electric_field_Fourier}
\end{equation}
where $V$ is the scattering region. The scattered wave's travel direction is expressed in spherical coordinates as $\bm{n}=(\sin \theta \cos \varphi, \sin \theta \sin \varphi, \cos \theta)$. The polar angle $\theta$ is defined as the angle between the $z$-axis direction and the direction of the scattered wave. Substituting equation \eqref{eq:motion_of_pm} into equation \eqref{eq:scattered_electric_field_Fourier}, we find
\begin{equation}
\begin{aligned}
&\widetilde{\bm{E}_{\text{tot}}}\left(\omega_1\right)=\frac{e^2 E_0}{c^2 R m_{\text{e}}} \frac{\omega_{0}^2}{\omega_{0}^2-\omega_{\mathrm{c}}^2}\int_{-\infty}^{\infty}\dd t \int_V \dd^3 \bm{r}~ e^{-i\omega_1\left(t^{\prime}+\frac{R}{c}-\frac{\bm{n} \cdot \bm{r}}{c}\right)}~\\
&\times\left[\left(\begin{array}{l}
\sin^2 \theta \sin \varphi \cos \varphi \\
-(1-\sin^2\theta\sin^2\varphi) \\
\sin\theta\cos\theta\sin\varphi
\end{array}\right) \cos \left(\bm{k}_{0}\cdot\bm{r}-\omega_0t\right)\right.\\
&\times\left\{ \delta n_{+}\left(\bm{r},t\right)+\delta n_{-}\left(\bm{r},t\right)\right\}-\frac{\omega_{\text{c}}}{\omega_0}\left(\begin{array}{l}
\sin\theta\cos\theta\cos\varphi \\
\sin\theta\cos\theta\sin\varphi \\
-\sin^2\theta
\end{array}\right)\\
&\left.\times\sin \left(\bm{k}_{0}\cdot\bm{r}-\omega_0t\right)\left\{\delta n_{+}\left(\bm{r},t\right)-\delta n_{-}\left(\bm{r},t\right)\right\}\right].
\end{aligned}    
\end{equation}
The density fluctuations of electrons and positrons are Fourier-transformed with respect to space and time as
\begin{equation}
\delta n_{ \pm}\left(\bm{r},t\right)=\frac{1}{(2 \pi)^4} \int \dd^3 \bm{k}~ \dd \omega~ e^{i(\bm{k} \cdot \bm{r}-\omega t)} \widetilde{\delta n}_{ \pm}(\bm{k}, \omega).
\end{equation}
Then the Fourier-transformed scattered electric field is expressed by
\begin{equation}
\begin{aligned}
&\left|\widetilde{\bm{E}_{\text {tot }}}\left(\omega_1\right)\right|^2
=  \left(\frac{e^2 E_0}{2 c^2 R m_{\text{e}}} \frac{\omega_{0}^2}{\omega_{0}^2-\omega_{\mathrm{c}}^2}\right)^2\\
&\times\left\{\left|\widetilde{\delta n_{+}}+\widetilde{\delta n_{-}}\right|^2\left(1-\sin ^2 \theta\sin ^2 \varphi\right)\right. \\
& \left.+\left(\frac{\omega_{\text{c}}}{\omega_0}\right)^2\left|\widetilde{\delta n_{+}}- \widetilde{\delta n_{-}}\right|^2\sin^2\theta\right\}.
\end{aligned}
\label{eq:total_electric_filed_abs}
\end{equation}
Here, the argument for the wave vector and frequency of the density fluctuations is described by the difference between the scattered waves $(\bm{k}_1=\omega_{1}\frac{\bm{n}}{c}, \omega_1)$ and incident waves as follows
\begin{equation}
 \widetilde{\delta n_{\pm}}=\widetilde{\delta n_{\pm}}(\bm{k}_{1}-\bm{k}_{0},\omega_1-\omega_0).
\end{equation}

In the next section, we evaluate the combinations of density fluctuations that characterize the magnitude of the scattered electric field.
\subsection{Spectral density functions}
Spectral density functions for density fluctuations are defined as a physical quantity that characterizes the intensity of plasma scattering. The following four types of spectral density functions characterize the scattering of $e^{\pm}$ plasma:
\begin{equation}
\begin{aligned}
& S_{ \pm \pm}(\boldsymbol{k}, \omega) \equiv \lim _{V, T \rightarrow \infty} \frac{\left\langle\left|\widetilde{\delta n_{ \pm}}(\boldsymbol{k}, \omega)\right|^2\right\rangle_{\text {ensemble }}}{V T n_{\text{e}}}, \\
& S_{ \pm \mp}(\boldsymbol{k}, \omega) \equiv \lim _{V, T \rightarrow \infty} \frac{\left\langle\widetilde{\delta n_{ \pm}}(\boldsymbol{k}, \omega){\widetilde{\delta n_{\mp}}}^*(\boldsymbol{k}, \omega)\right\rangle_{\text {ensemble }}}{V T n_{\text{e}}}.
\end{aligned}
\label{eq:Spectral_density_functions}
\end{equation}
Here $\langle\cdots\rangle_{\text {ensemble }}$ denotes taking the statistical mean according to the plasma distribution function, and 
\begin{equation}
n_{\text{e}}\equiv n_{0+}=n_{0-}     
\end{equation}
represents the electron or positron uniform density.

Using equations \eqref{eq:radiation_power_per_solid_angle}, \eqref{eq:total_electric_filed_abs}, and \eqref{eq:Spectral_density_functions}, the energy radiated by the plasma per unit time, unit solid angle, and unit frequency can be expressed by
\begin{equation}
\begin{aligned}
& \left\langle\frac{\dd P_{\text{s}} }{\dd \Omega\dd \omega_1}\right\rangle_{\text {ensemble }}=\frac{V n_{\text{e}}}{\pi} \frac{c R^2}{4 \pi} \left(\frac{e^2 E_0}{2 c^2 R m_{\text{e}}} \frac{\omega_{0}^2}{\omega_{0}^2-\omega_{\text{c}}^2}\right)^2\\
&\times\left[\left(S_{++}+S_{+-}+S_{-+}+S_{--}\right)\left(1-\sin ^2 \theta\sin ^2 \varphi\right)\right.\\
&\left. +\left(\frac{\omega_{\text{c}}}{\omega_0}\right)^2\left( S_{++}+S_{--}-S_{+-}-S_{-+}\right)\sin^2\theta\right].
\end{aligned}
\end{equation}
The total scattering cross-section for $2Vn_{\text{e}}$ particles in the scattering region $V$ is determined by dividing the scattering energy by the energy flux of the incident electromagnetic wave: 
\begin{equation}
\frac{\dd \sigma^{(2Vn_{\text{e}})}}{\dd\Omega\dd \omega_1}=\left\langle\frac{\dd P_{\text{s}}}{\dd\Omega\dd\omega_1}\right\rangle_{\text {ensemble }}\cdot \frac{8\pi}{cE_0^2}.
\label{eq:definition_of_differential_cross_section}
\end{equation}
The differential cross-section for scattering into $\dd\Omega$ and $\dd \omega_1$ by the $2Vn_{\text{e}}$ particles is then
\begin{equation}
\begin{aligned}
&\frac{\dd \sigma^{(2Vn_{\text{e}})}}{\dd\Omega\dd \omega_1}=Vn_{\text{e}} \frac{e^4}{2 \pi m_{\text{e}}^2 c^4}\left(\frac{\omega_0^2}{\omega_0^2-\omega_{\text{c}}^2}\right)^2\\
&\times\left[\left(S_{++}+S_{+-}+S_{-+}+S_{--}\right)\left(1-\sin ^2 \theta\sin ^2 \varphi\right)\right.\\
&\left. +\left(\frac{\omega_{\text{c}}}{\omega_0}\right)^2\left( S_{++}+S_{--}-S_{+-}-S_{-+}\right)\sin^2\theta\right].
\end{aligned}
\label{eq:Collective_cross_section}
\end{equation}
The argument of the spectral density function is
\begin{equation}
 S=S(\bm{k}_{1}-\bm{k}_{0},\omega_1-\omega_0).
\end{equation}

Equation \eqref{eq:Collective_cross_section} is a general expression describing Thomson scattering in $e^{\pm}$ plasma. The spectral density function is evaluated by taking the statistical mean of the plasma distribution function that gives the initial conditions of the position and velocity of each particle before scattering, as shown below. Given the appropriate initial plasma conditions, the scattering cross-section can be obtained, considering the correlation between particles. 

We derive the density fluctuations of $e^{\pm}$ plasma in the presence of a background magnetic field. The density fluctuations for ion-electron plasma in a magnetic field were derived by \citet{1960CaJPh..38.1114F,1960RSPSA.259...79D,1961PhRv..122.1663S}. The density fluctuations for the case where the constituent particles of the plasma are electrons and positrons can be obtained by replacing ion mass with electron mass. In the following, the derivation of the plasma density fluctuation is briefly described, and the detailed derivation is given in the Appendix \ref{Ap:derivation_of_density_fluctuation}. 

First, the Fourier transform of the density fluctuations for time must be replaced by the Laplace transform to incorporate the initial conditions of the particles before scattering into the equations. Assuming that $t = 0$ is the time when the incident electromagnetic wave first enters the scattering region, the Fourier-Laplace transform of the density fluctuation can be described by
\begin{equation}
\begin{aligned}
 \widetilde{\delta n_{ \pm}}(\bm{k}, \omega)&=\int_0^{\infty} \dd t~ e^{-i(\omega-i \varepsilon) t} \int \dd^3 \bm{r} ~\delta n_{ \pm}(\bm{r}, t) e^{i \boldsymbol{k} \cdot \bm{r}} \\
& =\int \dd^3 \bm{v} ~\widetilde{\delta F}_{ \pm}(\bm{k}, \bm{v}, \omega).
\end{aligned}
\label{eq:density_fluctuation}
\end{equation}
Here $\varepsilon$ is a positive infinitesimal quantity, a regularization factor that represents the elimination of effects due to scattering in the infinite future. From equation \eqref{eq:density_fluctuation}, the density fluctuations of electrons and positrons are represented by the first-order perturbation of the plasma's distribution function. This first-order perturbation can be obtained by using the Vlasov equation for the distribution function and Gauss's laws, with the non-perturbed components taken as the zeroth-order distribution functions $F_{0\pm}$, particle velocities $\bm{v}_{\pm}$, and the background magnetic field $\bm{B}_0$. The perturbed components are expressed by the first-order distribution functions $\delta F_{\pm}$ and the fluctuating electric field $\bm{E}$ generated by the plasma as follows
\begin{equation}
\begin{aligned}
& \frac{\partial F_{ 0\pm}}{\partial t}+\bm{v}_{ \pm} \cdot \frac{\partial F_{ 0\pm}}{\partial \bm{r}_{ \pm}} \pm \frac{e}{m_{\text{e}} c}\left(\bm{v}_{ \pm} \times \bm{B}_0\right) \cdot \frac{\partial F_{ 0\pm}}{\partial \bm{v}_{ \pm}}=0, \\
& \frac{\partial \delta F_{ \pm}}{\partial t}+\bm{v}_{ \pm} \cdot \frac{\partial \delta F_{ \pm}}{\partial \bm{r}_{ \pm}} \\
&\pm \frac{e}{m_{\text{e}} c}\left(\bm{v}_{ \pm} \times \bm{B}_0\right) \cdot \frac{\partial\delta F_{ \pm}}{\partial \bm{v}_{ \pm}} \pm \frac{e}{m_{\text{e}}} \bm{E} \cdot \frac{\partial F_{0 \pm}}{\partial \bm{v}_{ \pm}}=0, \\
& \nabla \cdot \bm{E}=4\pi\rho=\sum_{q= \pm e} 4 \pi q \int \dd^3 \bm{v}~ \delta F_{ \pm}.
\end{aligned}
\label{eq:Boltzmann_eq_for_magnetized_plasma}
\end{equation}
Let $f_{\pm}(\bm{v})$ be an one-particle distribution function, we can write $F_{0 \pm}\equiv n_{\text{e}}f_{\pm}(\bm{v})$. 

Before scattering, each plasma particle is in cyclotron motion in the background magnetic field. The initial velocity, position and phase of a particle are given by
\begin{equation}
\begin{array}{l}
\bm{v}_{ \pm}(t)=\left(\begin{array}{c}
v_{\parallel} \\
v_{\perp} \cos \varphi_{ \pm}(t) \\
v_{\perp} \sin \varphi_{ \pm}(t)
\end{array}\right), \\
\bm{r}_{ \pm}(t)=\bm{r}_{\pm}(0)+\left(\begin{array}{c}
v_{\parallel} t \\
\pm r_{\text{L}} \sin \varphi_{ \pm}(t) \\
\mp r_{\text{L}} \cos \varphi_{ \pm}(t)
\end{array}\right), \\
\varphi_{ \pm}(t) \equiv \pm \omega_{\text{c}} t+\phi_0,
\end{array}
\label{eq:initial_condition_for_plasma}
\end{equation}
where $v_{\|}$ and $v_{\perp}$ are the velocities in the direction parallel and perpendicular to the background magnetic field, $\phi_0$ is the angle between the particle position just before scattering and the $z$-axis, and 
\begin{equation}
r_{\text{L}}\equiv\frac{v_{\perp}}{\omega_{\text{c}}}    
\end{equation}
is so-called Lamor radius.
$\bm{r}_{\pm}(0)$ represents the particle position just before scattering, and each particle follows a canonical distribution in the electrostatic potential created by the plasma.

From equation \eqref{eq:Boltzmann_eq_for_magnetized_plasma}, we can evaluate the density fluctuations for $e^{\pm}$ plasma satisfying the initial condition \eqref{eq:initial_condition_for_plasma}. The density fluctuation for $e^{\pm}$ plasma in a background magnetic field can be written as follows, referring to the calculations of \citet{1961PhRv..122.1663S} who derived the density fluctuation for ion-electron plasma. The detailed derivation is given in Appendix \ref{Ap:derivation_of_density_fluctuation}
\begin{equation}
\begin{aligned}
& \widetilde{\delta n_{\pm}}(\boldsymbol{k}, \omega)=-i\left[\left(1-\frac{H_{\pm}}{\varepsilon_{\mathrm{L}}}\right)\sum_{j=1}^{N_{\pm}} e^{i \boldsymbol{k} \cdot \boldsymbol{r}_{\pm j}(0)}\right. \\
& \left. \times \sum_{l, m=-\infty}^{+\infty} \frac{J_l\left(\pm k_{\perp} r_{\text{L}}\right) J_m\left(\pm k_{\perp} r_{\text{L}}\right)}{\omega-i\varepsilon-k_x v_{\| }\mp l \omega_{\mathrm{c}}} e^{i(l-m) \phi_{0j}}+\frac{H_{\pm}}{\varepsilon_{\mathrm{L}}} \right. \\
& \left. \times\sum_{h=1}^{N_{\mp}} e^{i \boldsymbol{k} \cdot \boldsymbol{r}_{\mp h}(0)}\sum_{l, m=-\infty}^{+\infty} \frac{J_l\left(\mp k_{\perp} r_{\text{L}}\right) J_m\left(\mp k_{\perp} r_{\text{L}}\right)}{\omega-i\varepsilon-k_x v_{\| }\pm l \omega_{\mathrm{c}}} e^{i(l-m) \phi_{0h}}\right],
\end{aligned}
\label{eq:density_fluctuation_of_ep}
\end{equation}
where $J_l(z)$ is a Bessel function, and
\begin{equation}
    k_{\perp}\equiv\sqrt{k_y^2+k_z^2}.
\end{equation}
Here, $\varepsilon_{\text{L}}$ and $H_{\pm}$ are the longitudinal dielectric function and the positron/electron electric susceptibility, respectively, and can be described by
\begin{equation}
\begin{aligned}
& \varepsilon_{\text{L}}(\boldsymbol{k}, \omega) \equiv \frac{\bm{k} \cdot \bm{\varepsilon}(\boldsymbol{k}, \omega) \cdot \bm{k}}{k^2}, \\
& H_{ \pm}(\boldsymbol{k}, \omega) \equiv \frac{4 \pi i}{\omega} \frac{\bm{k} \cdot \bm{\sigma}_{\pm}(\boldsymbol{k}, \omega) \cdot \bm{k}}{k^2},
\end{aligned}
\label{eq:electric_susceptibility_and_longitudinal_dielectric_constant}
\end{equation}
where $\bm{\varepsilon}$ is the dielectric tensor of a magnetized plasma and $\bm{\sigma}_{\pm}$ is the positron/electron electrical conductivity tensor, which are related to each other by $\bm{\varepsilon}=\bm{I}+\frac{4\pi i}{\omega}(\bm{\sigma}_{+}+\bm{\sigma}_{-})$.

The density fluctuation equation \eqref{eq:density_fluctuation_of_ep} is divided into three terms, each with a physical interpretation. Specifically, we focus on the electron density fluctuations $\widetilde{\delta n_{-}}(\boldsymbol{k}, \omega)$: the first term without $H_{\pm}/\varepsilon_{\mathrm{L}}$ and summed over the electron indices is called the non-collective term, which means that each electron is in cyclotron motion in the background magnetic field; the second term with $H_{\pm}/\varepsilon_{\mathrm{L}}$ and summed over the electron indices is the effect of each electron on the rest of the electron population; the third term with $H_{\pm}/\varepsilon_{\mathrm{L}}$ and summed over the positron indices is the effect of each positron on the electron population. The second and third terms are called the collective term, which is the effect of the particles that make up the plasma being distributed and correlated.

Depending on whether the time variable of the density fluctuation is Fourier-transformed or Laplace-transformed, the expression of the spectral density function differs as
\begin{equation}
\begin{aligned}
    S(\boldsymbol{k}, \omega) &\equiv \lim _{\substack{\varepsilon\rightarrow 0 \\ V \rightarrow \infty}} \frac{2 \varepsilon}{V} \frac{\left\langle\left|\widetilde{\delta n}(\boldsymbol{k}, \omega)\right|^2\right\rangle_{\text {ensemble }}^{\text {Laplace }}}{n_{\text{e}}}\\
    &=\lim _{\substack{T \rightarrow \infty \\ V \rightarrow \infty}} \frac{1}{T V} \frac{\left\langle\left|\widetilde{\delta n}(\boldsymbol{k}, \omega)\right|^2\right\rangle_{\text {ensemble }}^{\text {Fourier }}}{n_{\text{e}}}.
\end{aligned}
\label{eq:spectral_density_function_laplace}
\end{equation}
Equation \eqref{eq:spectral_density_function_laplace} are equivalent to each other according to the ergodic hypothesis that taking a long-time average is equivalent to taking a multi-particle statistical average.

Using the expression for density fluctuations, four spectral density functions can be evaluated. The equation below is one example of substituting density fluctuations \eqref{eq:density_fluctuation_of_ep} into the definition of $S_{--}(\boldsymbol{k}, \omega)$ expressed as
\begin{equation}
\begin{aligned}
&S_{--}(\boldsymbol{k}, \omega)=\lim _{\substack{\varepsilon \rightarrow 0 \\ V \rightarrow \infty}} \frac{2 \varepsilon}{V n_{\text{e}}}\left\langle\left\{\left(1-\frac{H_{-}}{\varepsilon_{\text{L}}}\right)\right.\right.\\
& \left.\left.\times\sum_{j=1}^{N_{-}} e^{i \bm{k} \cdot \bm{r}_{-j}(0)} \sum_{l, m} \frac{J_l\left(-k_{\perp} r_{\text{L}}\right) J_m\left(-k_{\perp} r_{\text{L}}\right)}{\omega-k_x v_{\| }+l \omega_{\text{c}}-i \varepsilon} e^{i(l-m) \phi_{0j}}\right.\right.\\
&\left.+\frac{H_{-}}{\varepsilon_{\text{L}}} \sum_{h=1}^{N_{+}} e^{i \boldsymbol{k} \cdot \boldsymbol{r}_{+h}(0)} \sum_{l, m} \frac{J_l\left(k_{\perp} r_{\text{L}}\right) J_m\left(k_{\perp} r_{\text{L}}\right)}{\omega-k_x v_{\|}-l \omega_{\text{c}}-i \varepsilon} e^{i(l-m) \phi_{0h}}\right\}\\
&\left\{\left(1-\frac{H_{-}^{*}}{\varepsilon_{\text{L}}^{*}}\right) \sum_{s=1}^{N_{-}} e^{-i \bm{k} \cdot \bm{r}_{-s}(0)} \sum_{l^{\prime}, m^{\prime}} \frac{J_{l^{\prime}}\left(-k_{\perp} r_{\text{L}}\right) J_{m^{\prime}}\left(-k_{\perp} r_{\text{L}}\right)}{\omega-k_x v_{\| }+l^{\prime} \omega_{\text{c}}+i \varepsilon}\right.\\&\left.\times e^{-i(l^{\prime}-m^{\prime}) \phi_{0s}}+\frac{H_{-}^{*}}{\varepsilon_{\text{L}}^{*}} \sum_{g=1}^{N_{+}} e^{-i \boldsymbol{k} \cdot \boldsymbol{r}_{+g}(0)}\right.\\
&\times\left.\left. \sum_{l^{\prime}, m^{\prime}} \frac{J_{l^{\prime}}\left(k_{\perp} r_{\text{L}}\right) J_{m^{\prime}}\left(k_{\perp} r_{\text{L}}\right)}{\omega-k_x v_{\| }-l^{\prime}\omega_{\text{c}}+i \varepsilon} e^{-i(l^{\prime}-m^{\prime}) \phi_{0g}}\right\}\right\rangle.
\end{aligned}
\label{eq:spectral_density_function_mid}
\end{equation}
The spectral density function is a multiplication of terms describing the motion of each of the $N_{-}$ electrons and $N_{+}$ positrons in the scattering region. If we take a statistical average over a population of particles, only the product of the identical particles remains. That is, only terms with indices $j = s$ and $h = g$ remain. The physical interpretation is that the phase factor of equation \eqref{eq:spectral_density_function_mid}, $e^{i \bm{k} \cdot \bm{r}_{-j}(0)}$, completely cancels out only the terms of the identical particles, while the terms of different particles do not cancel out entirely and converge to zero when the statistical average is taken (see, e.g., \citep{1960PhRv..120.1528S,2012FuST...61..104F}). This is because the position of each particle before scattering is randomly distributed according to the canonical distribution. Furthermore, regarding the infinite sums of Bessel functions for $l$, $l^{\prime}$, $m$, and $m^{\prime}$, only the terms with $l=l^{\prime}$ and $m=m^{\prime}$ remain, and all other terms become zero when taking the statistical average.

The four spectral density functions are expressed in the following using the infinite sum formula for Bessel functions $\sum_{m=-\infty}^{+\infty} J_m^2(z)=1$,
\begin{equation}
\begin{aligned}
&S_{\pm\pm}(\boldsymbol{k}, \omega)
=  \lim _{\substack{\varepsilon \rightarrow 0 \\
V \rightarrow \infty}} 2 \varepsilon\\
\times&\left[\left|1-\frac{H_{\pm}}{\varepsilon_{\text{L}}}\right|^2 \sum_{l=-\infty}^\infty \int \dd^{3} \bm{v} \frac{J_l^2\left(\pm k_{\perp} r_{\text{L}}\right) f_{\pm}(\bm{v})}{\left(\omega-k_x v_{\|}\mp l \omega_{\text{c}}\right)^2+\varepsilon^2}\right. \\
& \left.+\left|\frac{H_{\pm}}{\varepsilon_{\text{L}}}\right|^{2} \sum_{l=-\infty}^{+\infty} \int \dd^3 \bm{v} \frac{J_l^2\left(\mp k_{\perp} r_{\text{L}}\right) f_{\mp}(\bm{v})}{\left(\omega-k_xv_{\|}\pm l \omega_{\text{c}}\right)^2+\varepsilon^2}\right],
\end{aligned}
\label{eq:Spectral_density_function_general1}
\end{equation}
\begin{equation}
\begin{aligned}
&S_{\pm\mp}(\boldsymbol{k}, \omega)
=  \lim _{\substack{\varepsilon \rightarrow 0 \\
V \rightarrow \infty}} 2 \varepsilon\\
\times&\left[\left(1-\frac{H_{\pm}}{\varepsilon_{\text{L}}}\right)\frac{H_{\mp}^{*}}{\varepsilon_{\text{L}}^{*}} \sum_{l=-\infty}^\infty \int \dd^{3} \bm{v} \frac{J_l^2\left(\pm k_{\perp} r_{\text{L}}\right) f_{\pm}(\bm{v})}{\left(\omega-k_x v_{\|}\mp l \omega_{\text{c}}\right)^2+\varepsilon^2}\right. \\
& \left.+\left(1-\frac{H_{\mp}^{*}}{\varepsilon_{\text{L}}^{*}}\right)\frac{H_{\pm}}{\varepsilon_{\text{L}}} \sum_{l=-\infty}^{+\infty} \int \dd^3 \bm{v} \frac{J_l^2\left(\mp k_{\perp} r_{\text{L}}\right) f_{\mp}(\bm{v})}{\left(\omega-k_xv_{\|}\mp l \omega_{\text{c}}\right)^2+\varepsilon^2}\right].
\end{aligned}
\label{eq:Spectral_density_function_general2}
\end{equation}
The spectral density function and Thomson cross-section for $e^{\pm}$ plasma in a background magnetic field can be obtained from these expressions. In the next section, we assume the Maxwellian distribution for the plasma distribution function and discuss the behavior of plasma scattering.
\subsection{Maxwellian distributions}
In this section, we investigate the behavior of Thomson scattering in the case of a Maxwellian distribution of $e^{\pm}$ plasma. First, we evaluate the electric susceptibility and longitudinal dielectric function in magnetized electron and positron plasma that appear in the spectral density functions. According to the plasma kinetic theory (e.g., \citep{2012FuST...61..104F}), the expression for the electric susceptibility $H_{ \pm}$, \eqref{eq:electric_susceptibility_and_longitudinal_dielectric_constant}, is given by

\begin{equation}
\begin{aligned}
H_{ \pm}(\bm{k}, \omega)=&\int \dd^3 \bm{v} \frac{4 \pi e^2 n_{\text{e}}}{m_{\text{e}} k^2} \sum_{l=-\infty}^{+\infty}\left(k_x \frac{\partial f_{0\pm}}{\partial v_x} \pm \frac{l \omega_{\text{c}}}{v_{\perp}} \frac{\partial f_{0 \pm}}{\partial v_{\perp}}\right) \\&\times\frac{J_l^2\left( \pm \frac{k_{\perp} v_{\perp}}{\omega_{\text{c}}}\right)}{\omega-i \varepsilon-k_x v_x \mp l \omega_{\text{c}}}.
\end{aligned}
\label{eq:electric_susceptibility_thermal}
\end{equation}
Assume the following Maxwellian distribution as the $e^{\pm}$ plasma distribution function
\begin{equation}
\begin{aligned}
f_{ \pm}(\bm{v}) & =\left(\frac{m_{\text{e}}}{2 \pi k_{\text{B}} T_{\text{e}}}\right)^{\frac{3}{2}} \exp \left(-\frac{m_{\text{e}} v^2}{2 k_{\text{B}} T_{\text{e}}}\right) \\
& =\frac{1}{\left(\pi v_{\text{th}}^2\right)^{\frac{3}{2}}} \exp \left(-\frac{v_x^2+v_{\perp}^2}{v_{\text{th}}^2}\right),
\end{aligned}
\label{eq:Maxwellian_distribution}
\end{equation}
where the thermal velocity is denoted by
\begin{equation}
v_{\text{th}}\equiv\left(\frac{2 k_{\text{B}} T_{\text{e}}}{m_{\text{e}}}\right)^{\frac{1}{2}}    
\end{equation}
and the thermal velocity in the direction parallel and perpendicular to the magnetic field is assumed to be equal. Substituting the Maxwellian distribution into equation \eqref{eq:electric_susceptibility_thermal} and performing velocity integration, the electric susceptibility can be represented by the modified Bessel function $I_l(x)$ and plasma dispersion function $Z(x)$. Using the following special function formulae,
\begin{eqnarray}
\int_0^{\infty}&& J_l^2(a t) \exp \left(-b^2 t^2\right) t ~\dd t \nonumber\\
&=&\frac{1}{2 b^2} \exp \left[-\left(\frac{a^2}{2 b^2}\right)\right] I_l\left(\frac{a^2}{2 b^2}\right),  \label{eq:integral_formula_modified_bessel}\\
Z(\xi) &\equiv& \frac{1}{\sqrt{\pi}} \int_{-\infty}^{\infty} \frac{1}{z-\xi} e^{-z^2} \mathrm{~d} z,
\end{eqnarray}
in performing the velocity integral of the electric susceptibility, it can be written as
\begin{equation}
\begin{aligned}
H_{ \pm}(\bm{k}, \omega)&=\frac{\omega_{\text{p}}^2}{k^2 v_{\text{th}}^2}\left\{1+\frac{\omega}{k_x v_{\text{th}}} \sum_l I_l\left[\frac{1}{2}\left(\frac{k_{\perp} v_{\text{th}}}{\omega_{\text{c}}}\right)^2\right]\right.\\
&\left.\times\exp \left[-\frac{1}{2}\left(\frac{k_{\perp} v_{\text{th}}}{\omega_{\text{c}}}\right)^2\right] Z\left(\frac{\omega_{\mp} l \omega_{\text{c}}}{k_x v_{\text{th}}}-i \varepsilon\right)\right\},
\end{aligned}
\label{eq:electric_susceptibility_Maxwellian}
\end{equation}
where 
\begin{equation}
    \omega_{\text{p}}\equiv\sqrt{\frac{4 \pi \cdot 2 n_{\mathrm{e}} e^2}{m_{\mathrm{e}}}}
\end{equation}
is defined as the $e^{\pm}$ plasma frequency. From the symmetry $I_l(x)=I_{-l}(x)$ of the modified Bessel function, we see that the electric susceptibility of electrons and positrons in the Maxwellian distribution is equal
\begin{equation}
H_{+}(\bm{k}, \omega)=H_{-}(\bm{k}, \omega)\equiv H(\bm{k}, \omega).   
\label{eq:susceptibility_sym}
\end{equation}

Next, we evaluate four types of spectral density functions \eqref{eq:Spectral_density_function_general1} and \eqref{eq:Spectral_density_function_general2} under the assumption of the Maxwellian distribution. Using the formula \eqref{eq:integral_formula_modified_bessel}, the integral appearing in the spectral density functions is calculated as 
\begin{equation}
\begin{aligned}
&\lim _{\varepsilon \rightarrow 0} 2 \varepsilon \sum_{l=-\infty}^{+\infty} \int \dd^3 \bm{v} \frac{J_l^2\left(\pm k_{\perp} r_{\text{L}}\right) f_{ \pm}(\bm{v})}{\left(\omega-k_{x} v_{x} \mp l \omega_{\text{c}}\right)^2+\varepsilon^2}    \\
=&\lim _{\varepsilon \rightarrow 0} 2 \varepsilon \sum_l  \int_0^{2 \pi} \frac{\dd \varphi}{\left(\pi v_{\text{th}}^2\right)^{\frac{3}{2}}} \int_0^{\infty} v_{\perp} \dd v_{\perp} J_l^2\left(k_{\perp} r_{\text{L}}\right) \exp \left(-\frac{v_{\perp}^2}{v_{\text{th}}^2}\right) \\
& \times \int_{-\infty}^{+\infty} \dd v_{x} \frac{\exp \left(-v_x^2 / v_{\text{th}}^2\right)}{\left(\omega-k_{x} v_{x}\mp l\omega_{\text{c}}\right)^2+\varepsilon^2}\\
=&~2 \sqrt{\pi} \sum_{l=-\infty}^{+\infty} \exp \left\{-\frac{1}{2}\left(\frac{v_{\text{th}} k_{\perp}}{\omega_{\text{c}}}\right)^2\right\} I_l\left[\frac{1}{2}\left(\frac{v_{\text{th}} k_{\perp}}{\omega_{\text{c}}}\right)^2\right]\\
&\times\frac{\exp \left[-\left(\frac{\omega \mp l \omega_{\text{c}}}{k_x v_{\text{th}}}\right)^2\right]}{k_x v_{\text{th}}}.
\end{aligned}
\end{equation}
From the symmetry of the modified Bessel function $I_l(x)=I_{-l}(x)$ and the equation \eqref{eq:susceptibility_sym}, we obtain the spectral density function as 
\begin{equation}
\begin{aligned}
S_{++}&=S_{--}=2 \sqrt{\pi}  \left(\left|1-\frac{H}{\varepsilon_{\text{L}}}\right|^2+\left|\frac{H}{\varepsilon_{\text{L}}}\right|^2\right) \\
& \times \sum_{l=-\infty}^{+\infty} \exp \left\{-\frac{1}{2}\left(\frac{v_{\text{th}} k_{\perp}}{\omega_{\text{c}}}\right)^2\right\} I_l\left[\frac{1}{2}\left(\frac{v_{\text{th}} k_{\perp}}{\omega_{\text{c}}}\right)^2\right]\\
&\times\frac{\exp \left[-\left(\frac{\omega - l \omega_{\text{c}}}{k_x v_{\text{th}}}\right)^2\right]}{k_x v_{\text{th}}},
\end{aligned}
\end{equation}
\begin{equation}
\begin{aligned}
S_{+-}&=S_{-+}=2 \sqrt{\pi}  \left\{\left(1-\frac{H}{\varepsilon_{\text{L}}}\right) \frac{H^*}{\varepsilon_{\text{L}}^*}+\left(1-\frac{H^*}{\varepsilon_{\text{L}}^*}\right) \frac{H}{\varepsilon_{\text{L}}}\right\} \\
& \times \sum_{l=-\infty}^{+\infty} \exp \left\{-\frac{1}{2}\left(\frac{v_{\text{th}} k_{\perp}}{\omega_{\text{c}}}\right)^2\right\} I_l\left[\frac{1}{2}\left(\frac{v_{\text{th}} k_{\perp}}{\omega_{\text{c}}}\right)^2\right]\\
&\times\frac{\exp \left[-\left(\frac{\omega - l \omega_{\text{c}}}{k_x v_{\text{th}}}\right)^2\right]}{k_x v_{\text{th}}}.
\end{aligned}
\end{equation}
Hence, the linear combinations of the spectral density functions appearing in the differential scattering cross-section \eqref{eq:Collective_cross_section} can be obtained as follows
\begin{equation}
\begin{aligned}
& S_{++}+S_{--}+S_{-+}+S_{+-} \\
= & 4 \sqrt{\pi} \sum_{l=-\infty}^{+\infty} \exp \left\{-\frac{1}{2}\left(\frac{v_{\text{th}} k_{\perp}}{\omega_{\text{c}}}\right)^2\right\} I_l\left[\frac{1}{2}\left(\frac{v_{\text{th}} k_{\perp}}{\omega_{\text{c}}}\right)^2\right]\\
&\times\frac{\exp \left[-\left(\frac{\omega - l \omega_{\text{c}}}{k_x v_{\text{th}}}\right)^2\right]}{k_x v_{\text{th}}},
\end{aligned}
\label{eq:spectral_density_function_electric}
\end{equation}
\begin{equation}
\begin{aligned}
& S_{++}+S_{--}-S_{-+}-S_{+-} \\
& =4 \sqrt{\pi}\left\{1-4 \operatorname{Re}\left(\frac{H}{\varepsilon_{\text{L}}}\right)+4\left|\frac{H}{\varepsilon_{\text{L}}}\right|^2\right\} \\
&\times \sum_{l=-\infty}^{+\infty} \exp \left\{-\frac{1}{2}\left(\frac{v_{\text{th}} k_{\perp}}{\omega_{\text{c}}}\right)^2\right\} I_l\left[\frac{1}{2}\left(\frac{v_{\text{th}} k_{\perp}}{\omega_{\text{c}}}\right)^2\right]\\
&\times\frac{\exp \left[-\left(\frac{\omega - l \omega_{\text{c}}}{k_x v_{\text{th}}}\right)^2\right]}{k_x v_{\text{th}}}.
\end{aligned}
\label{eq:spectral_density_function_drift}
\end{equation}


\begin{table}[htbp]
  \centering
  \captionsetup{
    skip=1em,          
    justification=raggedright 
  }
  \caption{Parameters used for the differential cross-sections plotted at two different temperatures.}
  \label{tab:Collective_parameter}
  \renewcommand{\arraystretch}{1.5}
  \sisetup{table-number-alignment=left}
  \begin{tabular}{lSS}
    \toprule
    Parameter & {Value1} & {Value2} \\
    \midrule
    $\omega_0~[\si{\giga\hertz}/2\pi]$ & 1 & \\
    $\omega_{\text{c}}~[\si{\giga\hertz}/2\pi]$ & 2 & \\
    $n_{\text{e}}~[\si{\centi\meter^{-3}}]$ & {$6\times10^5$} & \\
    $T_{\text{e}}~[\si{\kilo\electronvolt}]$ & 153 & 5 \\
    \bottomrule
  \end{tabular}
\end{table}

\begin{figure*}
  \centering
  \begin{minipage}[b]{0.5\textwidth}
    \centering
    \textbf{(a)}
    \includegraphics[width=\textwidth,clip]{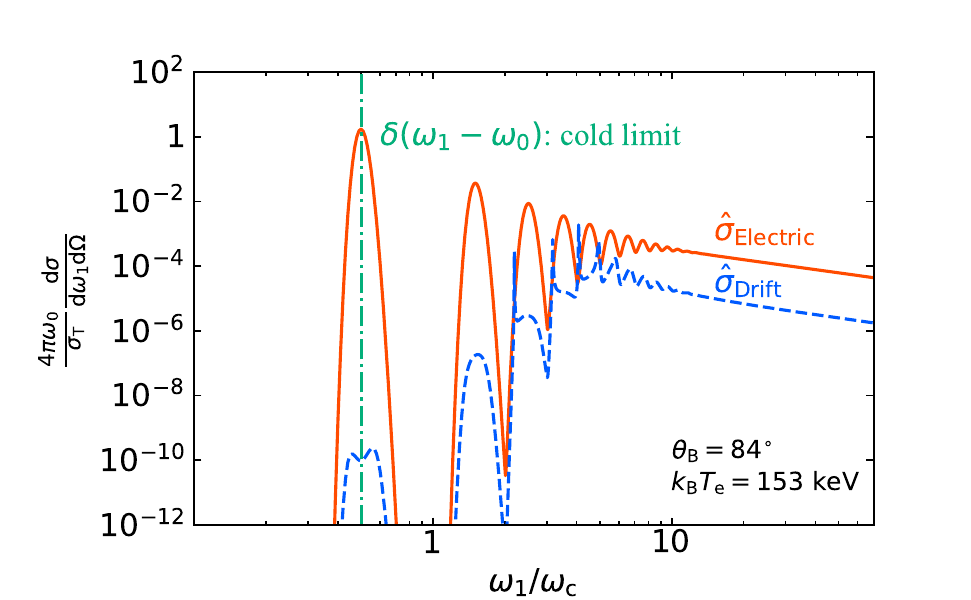}
    \label{fig_Collective_perp_high_temp}
  \end{minipage}%
  \begin{minipage}[b]{0.5\textwidth}
    \centering
    \textbf{(b)}
    \includegraphics[width=\textwidth,clip]{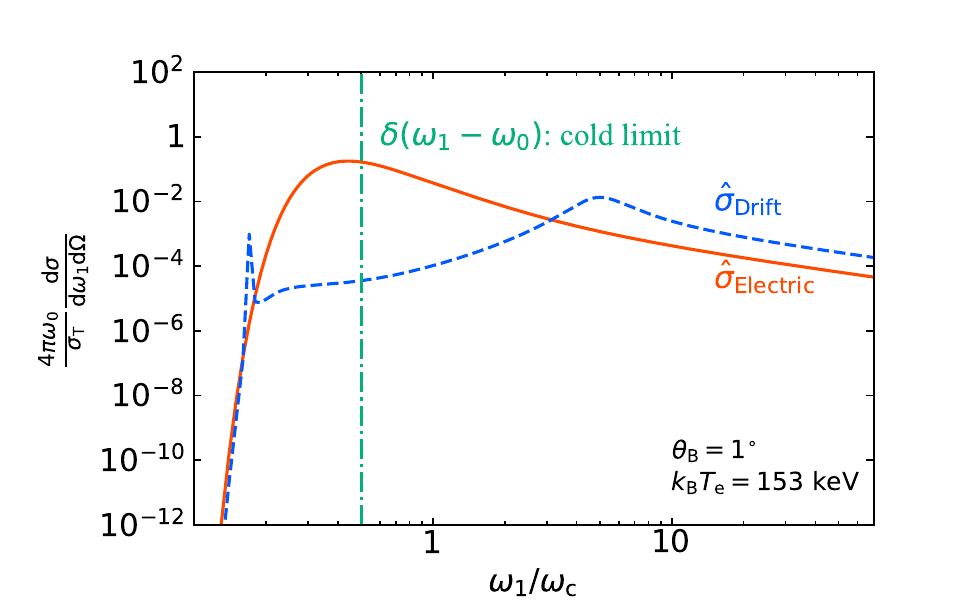}
    \label{fig_Collective_para_high_temp}
  \end{minipage}

  \vspace{\baselineskip}

  \begin{minipage}[b]{0.5\textwidth}
    \centering
    \textbf{(c)}
    \includegraphics[width=\textwidth,clip]{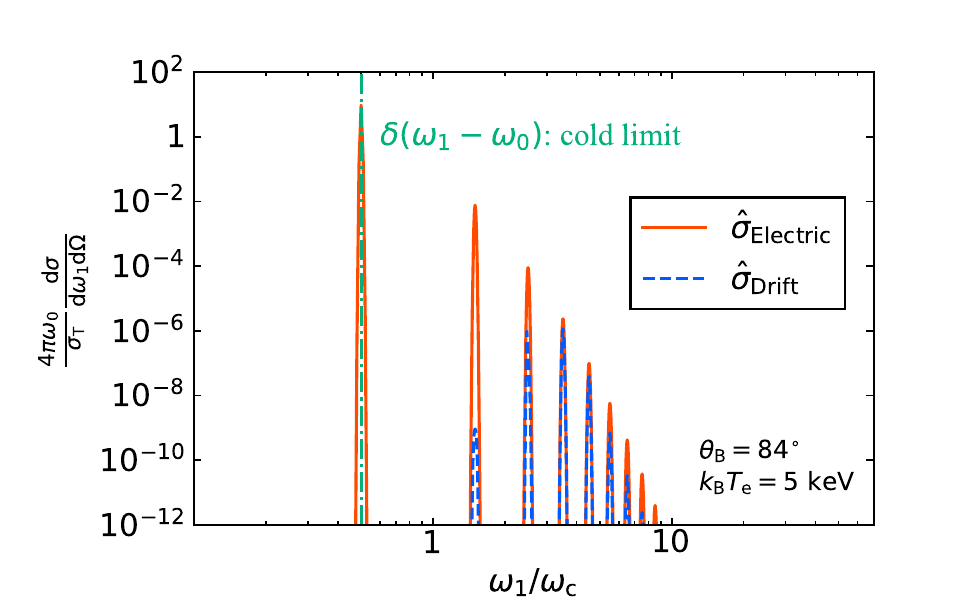}
    \label{fig_Collective_perp_low_temp}
  \end{minipage}%
  \begin{minipage}[b]{0.5\textwidth}
    \centering
    \textbf{(d)}
    \includegraphics[width=\textwidth,clip]{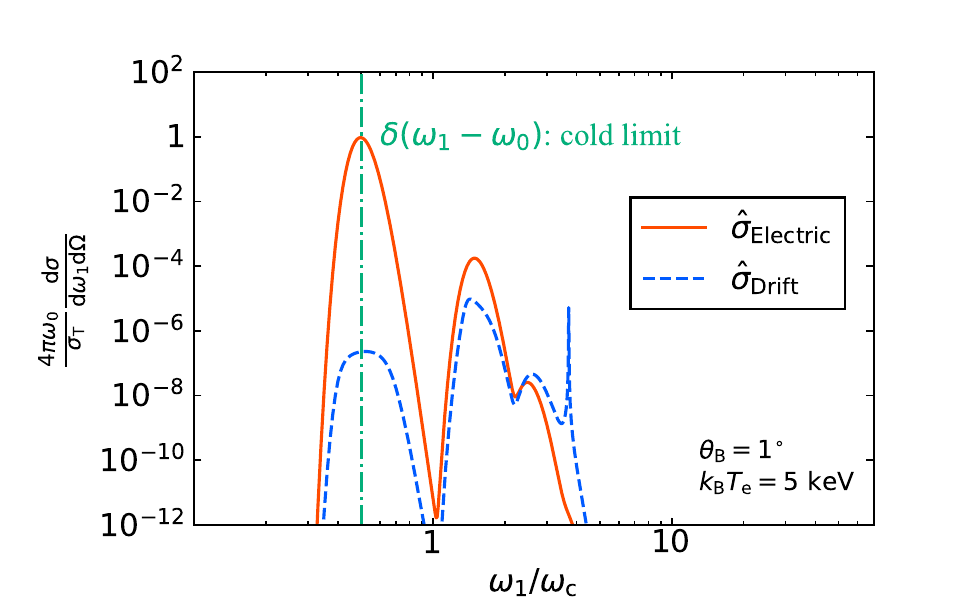}
    \label{fig_Collective_para_low_temp}
  \end{minipage}
    \captionsetup{
    skip=1em,          
    justification=raggedright 
  }
  \caption{Thomson scattering spectra from electron and positron plasma in a magnetic field. The solid red line represents the contribution from particles oscillating in the direction of the electric field of the incident electromagnetic wave, and the blue dashed line represents the contribution from the drift motion of the particles. The green dot-dashed line represents the spectrum when the cold plasma limit is taken~(see section \ref{subsec:cold_limit_of_collective_scattering} for details). The analysis is conducted using the parameters listed in Table \ref{tab:Collective_parameter}. (a) $k_{\text{B}}T_{\text{e}}=153~\text{keV}$ and the direction of the scattered wave is almost perpendicular to the magnetic field. (b) $k_{\text{B}}T_{\text{e}}=153~\text{keV}$ and the direction of the scattered wave is almost parallel to the magnetic field. (c) $k_{\text{B}}T_{\text{e}}=5~\text{keV}$ and the direction of the scattered wave is almost perpendicular to the magnetic field. (d) $k_{\text{B}}T_{\text{e}}=5~\text{keV}$ and the direction of the scattered wave is almost parallel to the magnetic field. The magnetic field value (cyclotron frequency) was chosen to show the spectrum clearly. Note that actual magnetic field strength strongly depends on the radius from the magnetar center. On a magnetar surface, the cyclotron frequency is several orders higher than radio frequencies. This leads to a scattering spectrum where only the peak at $\omega_0$ is pronounced, and higher frequency peaks beyond the second become less noticeable.}
  \label{fig_Collective_spectra}
\end{figure*}
Concerning equation \eqref{eq:spectral_density_function_electric}, the linear combination of the spectral density functions due to the motion of the oscillating particles in the incident electromagnetic wave direction is independent of the electric susceptibility $H$ and the scattering is strictly non-collective. On the other hand, in equation \eqref{eq:spectral_density_function_drift}, the term due to drift motion is dependent on the electric susceptibility $H$, and scattering has a collective effect.

In Figure \ref{fig_Collective_spectra}, we plot the spectra of the differential scattering cross-section \eqref{eq:Collective_cross_section} at two different temperature $T_{\text{e}}=153~\text{keV}$ and $T_{\text{e}}=5~\text{keV}$. The other parameters we use are shown in Table \ref{tab:Collective_parameter}. The differential cross-section is non-dimensionalized by the Thomson cross-section $\sigma_{\text{T}}$ and the angular frequency of the incident electromagnetic wave $\omega_0$ as
\begin{equation}
\begin{aligned}
& \frac{4\pi\omega_0}{\sigma_{\text{T}}}\frac{\mathrm{d} \sigma^{\left(1\right)}}{\mathrm{d} \Omega \mathrm{d} \omega_1}=\frac{3\omega_0}{8\pi}\left(\frac{\omega_0^2}{\omega_0^2-\omega_{\mathrm{c}}^2}\right)^2 \\
& \times\left[\left(S_{++}+S_{+-}+S_{-+}+S_{--}\right)\left(1-\sin ^2 \theta \sin ^2 \varphi\right)\right. \\
& \left.+\left(\frac{\omega_{\mathrm{c}}}{\omega_0}\right)^2\left(S_{++}+S_{--}-S_{+-}-S_{-+}\right)\sin^2\theta\right]\\
&\equiv\hat{\sigma}_{\text{Electric}}+\hat{\sigma}_{\text{Drift}},
\end{aligned}
\end{equation}
where the contribution due to oscillation of particles in the direction of electric field of the incident electromagnetic wave is defined as $\hat{\sigma}_{\text{Electric}}$ and the contribution due to drift motion is defined as $\hat{\sigma}_{\text{Drift}}$. We specifically selected the incident direction of the electromagnetic wave to be perpendicular to the background magnetic field. Additionally, we define the angle between the magnetic field and the propagation direction of the scattered wave as \(\theta_{\text{B}} \equiv \pi/2 - \theta\). Furthermore, we consider the case where the wave number vector of the scattered wave lies within the plane formed by the incident electromagnetic wave and the magnetic field (\(\varphi=0\)).  Figure \ref{fig_Collective_spectra} shows the differential cross-sections at two plasma temperatures for the cases where the scattering wave is nearly perpendicular to the background magnetic field ($\theta_{\text{B}} = 84^{\circ}$) and nearly parallel to it ($\theta_{\text{B}} = 1^{\circ}$)\footnote{The reason for not choosing the scattered wave to be completely perpendicular or parallel is that it would result in the differential cross-sections having delta-function-like peaks.}. 

First, in $\hat{\sigma}_{\text{Electric}}$, when the direction of the scattered wave is perpendicular to the background magnetic field (see the left side of Figure \ref{fig_Collective_spectra}), the first significant peak appears at the angular frequency $\omega_0$ of the incident electromagnetic wave and the following peaks at frequencies $\omega_0+n\omega_{\text{c}}$~($n$ is a natural number) decrease the height as $n$ becomes large. Moreover, as the temperature decreases, the peak width and height on the high-frequency side become smaller. As will be shown later, Thomson scattering spectrum in the cold plasma limit has a delta-function peak localized at the frequency $\omega_0$ of the incident electromagnetic wave. This corresponds to the case where single-particle scattering is considered. The peaks separated by the cyclotron frequency arise for finite temperature plasma because particles in the plasma, initially moving at integer multiples of the cyclotron frequency, are excited by the incident electromagnetic wave as density fluctuations. As the thermal motion of the particles becomes smaller, the peaks on the high-frequency side are suppressed, and the $\omega_0$ peaks become dominant.

Next, in $\hat{\sigma}_{\text{Electric}}$, when the scattered wave propagates parallel to the background magnetic field (see left side of Figure \ref{fig_Collective_spectra}), the peaks separated by the cyclotron frequency become less distinguishable. As the temperature decreases, the intensity on the high-frequency side diminishes.

Finally, for $\hat{\sigma}_{\text{Drift}}$, the rough configuration is the same as for $\hat{\sigma}_{\text{Electric}}$, but it has a more complicated shape due to its dependence on the electric susceptibility. The spectrum shows a spiky shape near the zero point of the longitudinal dielectric function ~($\varepsilon_{\text{L}}=0$), i.e., near the eigenmodes of electron and positron plasma in a magnetic field. The dependence of the scattering spectrum on the electric susceptibility, i.e., the collective effect, is suppressed at higher frequencies. This is because the electric susceptibility has a wavelength dependence \begin{equation}
    H(\bm{k},\omega)\propto\frac{\omega_{\text{p}}^2}{k^2 v_{\text{th}}^2} \sim\left(\frac{\lambda}{\lambda_{\text{De}}}\right)^2,
    \label{eq:Debye_dependence_of_electric_susceptibility}
\end{equation}
from equation \eqref{eq:electric_susceptibility_Maxwellian}, and becomes smaller at shorter wavelengths, i.e., higher frequencies. Here 
\begin{equation}
\lambda_{\text{De}}\equiv\left(\frac{k_{\text{B}}T_{\text{e}}}{8 \pi e^2 n_{\text{e}}}\right)^{\frac{1}{2}}=\frac{v_{\text{th}}}{\sqrt{2} \omega_{\text{p}}}
\end{equation}
is Debye length.

\subsection{Cold plasma limit}
\label{subsec:cold_limit_of_collective_scattering}
In this section, we estimate the spectral density function and calculate the scattering cross-section per particle in the limit where the thermal motion of plasma particles is negligible, and the background magnetic field is large. First, we evaluate the part of the spectral density function that depends on the electric susceptibility $H$. The electric susceptibility \eqref{eq:electric_susceptibility_Maxwellian} is expanded using Taylor series around $k_{\perp} v_{\text{th}}/\omega_{\text{c}}$ and $k_x v_{\text{th}}/(\omega\mp l\omega_{\text{c}})$, assuming that the thermal velocity is sufficiently small. The infinite sum of the modified Bessel functions converges quickly enough in $k_{\perp} v_{\text{th}}/\omega_{\text{c}}\ll 1$. Then the main orders of the electric susceptibility $H$ and the longitudinal dielectric function $\varepsilon_{\text{L}}=1+2H$ appear only for $l=-1,0,1$ and are expressed as
\begin{equation}
\begin{aligned}
   H(\bm{k},\omega) =&-\frac{1}{2} \frac{k_x^2}{k^2}\left(\frac{\omega_{\text{p}}}{\omega}\right)^2+\frac{1}{2} \frac{k_{\perp}^2}{k^2} \frac{\omega_{\text{p}}^2}{\omega_{\text{c}}^2}\left(1-\frac{\omega^2}{\omega^2-\omega_{\text{c}}^2}\right)\\
   &+\mathcal{O}\left(\left(\frac{kv_{\text{th}}}{\omega}\right)\right). 
\end{aligned}
\end{equation}
Then, the following Taylor expansion can be made on the order of cyclotron frequency
\begin{equation}
    \left.\frac{H_{\pm}}{\varepsilon_{\text{L}}}\right|_{\omega_1-\omega_0}=1+\mathcal{O}\left(\left(\frac{\omega_1-\omega_0}{\omega_{\text{c}}}\right)^2\right).
\end{equation}
Next, we evaluate the linear couplings \eqref{eq:spectral_density_function_electric} and \eqref{eq:spectral_density_function_drift} of the spectral density functions appearing in the differential cross-section \eqref{eq:Collective_cross_section}. If the angular frequency of the scattered wave is sufficiently smaller than the cyclotron frequency, i.e., if the magnetic field is sufficiently strong, only the term $l=0$ needs to be considered. Furthermore, in the cold plasma limit, the approximation $\frac{v_{\text{th}} k_{\perp}}{\omega_{\text{c}}} \ll 1$ holds. Given the above approximations, the infinite sum part of the spectral density function is evaluated as
\begin{equation}
\begin{aligned}
& \sum_{l=-\infty}^{+\infty} \exp \left\{-\frac{1}{2}\left(\frac{v_{\text{th}} k_{\perp}}{\omega_{\text{c}}}\right)^2\right\} I_l\left[\frac{1}{2}\left(\frac{v_{\text{th}} k_{\perp}}{\omega_{\text{c}}}\right)^2\right]\\
&\times\frac{\exp \left[-\left(\frac{\omega - l \omega_{\text{c}}}{k_x v_{\text{th}}}\right)^2\right]}{k_x v_{\text{th}}}\sim2 \sqrt{\pi}\frac{\exp \left[-\left(\frac{\omega}{k_x v_{\mathrm{th}}}\right)^2\right]}{k_x v_{\mathrm{th}}}.
\end{aligned}
\end{equation}
Finally, the sums of the spectral density functions are represented by the following delta function in the cold plasma limit
\begin{equation}
\begin{aligned}
&(\left.S_{++}+S_{--}+S_{+-}+S_{-+})\right|_{k_x, k_y, k_z-k_0, \omega-\omega_0}\\
\sim&4 \sqrt{\pi}\frac{\exp \left[-\left(\frac{\omega-\omega_0}{k_x v_{\mathrm{th}}}\right)^2\right]}{k_x v_{\mathrm{th}}}
\xrightarrow{v_{\mathrm{th}}\rightarrow0}4\pi\delta(\omega-\omega_0),
\end{aligned}
\label{eq:total_spectral_density_functions1}
\end{equation}
\begin{equation}
\begin{aligned}
&(\left.S_{++}+S_{--}-S_{+-}-S_{-+})\right|_{k_x, k_y, k_z-k_0, \omega-\omega_0}\\
&\xrightarrow{v_{\mathrm{th}}\rightarrow0}4\pi\delta(\omega-\omega_0).
\end{aligned}
\label{eq:total_spectral_density_functions2}
\end{equation}
The physical meaning of the spectral density function in the cold plasma limit is that the thermal motion of the plasma is sufficiently small that the spectrum of the scattered wave is localized to the frequency of the incident electromagnetic wave.

By substituting equations \eqref{eq:spectral_density_function_electric} and \eqref{eq:spectral_density_function_drift} into equation \eqref{eq:Collective_cross_section}, we obtain the cold plasma limit of the differential cross-section per particle.
\begin{equation}
\begin{aligned}
& \frac{\mathrm{d} \sigma_{\text{cold}}^{(1)}}{\mathrm{d} \Omega \mathrm{d} \omega_1}=r_{\mathrm{e}}^2\left[\left(\frac{\omega_0^2}{\omega_0^2-\omega_{\mathrm{c}}^2}\right)^2\left(1-\sin ^2 \theta \sin ^2 \varphi\right)\right. \\
& \left.+\left(\frac{\omega_0 \omega_{\mathrm{c}}}{\omega_0^2-\omega_{\mathrm{c}}^2}\right)^2\sin^2\theta\right]\delta(\omega_{1}-\omega_0).
\end{aligned}
\end{equation}
Then, the total cross-section per particle can be obtained by integrating over the frequency and solid angle of the scattered wave
\begin{equation}
\sigma_{\text{cold}}^{(1)}=\frac{1}{2} \sigma_{\mathrm{T}}\left[\left(\frac{\omega_0}{\omega_0+\omega_{\mathrm{c}}}\right)^2+\left(\frac{\omega_0}{\omega_0-\omega_{\mathrm{c}}}\right)^2\right].
\label{eq:collective_total_cross_section_cold}
\end{equation}
This is the same as when considering single particle scattering \eqref{eq:one_particle_cross_section}.

\section{DISCUSSION}
\label{sec:discussion_collective}
\subsection{Difference from the case without background magnetic fields}
Thomson scattering in unmagnetized $e^{\pm}$ plasma has been studied by \citet{1992PhFlB...4.2669S}. Their research concluded that the collective effects completely cancel out. Consider the radiation emitted by an electron oscillated by an external electric field. Around this electron are clouds of electrons and positrons, which attempt to neutralize the radiation of this electron due to the effect of Debye shielding. The electron cloud emits radiation in the opposite phase to the electron, and the positron cloud attempts to do the same\footnote{When considering acoustic waves, charged particle clouds oscillate irrespective of the charge sign. However, acoustic waves are hardly excited within the linear density fluctuations in $e^{\pm}$ plasma. Therefore, this study will not address the inherent oscillations of acoustic waves.}. The crucial point is that the accelerations of the electron and positron, when oscillated by the external electric field, are in opposite phases. Due to the Lienard-Wiechert potential, where the electric field from radiation is proportional to the acceleration of the radiating particles \( \bm{\dot{\beta}} \), these opposing accelerations cause the radiation of the positron cloud to be in the same phase as the electron. As a result, the radiations from the electron and positron clouds cancel each other out. Thus, the effect of Debye shielding does not appear in the radiation from the electron undergoing acceleration motion in $e^{\pm}$ plasma. The same argument applies to the radiation from positrons \footnote{Even when incorporating the electric field produced by the longitudinal eigenmode into the motion equation of the scattered particles \eqref{eq:EOM_for_plasma}, the cancellation of the collective effect is guaranteed. This holds especially when considering the nonlinear ponderomotive force arising from coupling the incident electric field and the eigenmode. The ponderomotive force acts similarly on both electrons and positrons, leading to a cancellation within the pair plasma. As a result, it does not generate an electric field in the plasma and is believed not to contribute to the radiation.}.

On the other hand, when a magnetic field is present, the collective effects in the scattering within $e^{\pm}$ plasma do not entirely cancel out. From equation \eqref{eq:spectral_density_function_electric}, it can be seen that the density fluctuations due to the motion of scattering particles in the direction of the incident electric field are entirely canceled by the collective effects, similar to the case without a background magnetic field. However, according to equation \eqref{eq:spectral_density_function_drift}, the density fluctuations due to drift motion direction retain some collective effects. This is because the acceleration of the radiating particle \( \bm{\dot{\beta}} \) has both a component dependent on the charge sign in the direction of the incident electric field and a component independent of the charge sign in the drift motion direction (see equations \eqref{eq:velocity_of_moving_particles}). Thus, the radiation originating from the drift does not cancel out.

In any case, as shown in the previous section, in strongly magnetized cold plasma limit, the scattering cross-section per particle is equal to that of the case of single particle scattering~(see equation\eqref{eq:one_particle_cross_section} and \eqref{eq:collective_total_cross_section_cold}), i.e., the collective effect can be neglected. The reason why the collective effect is negligible can be interpreted by focusing on the behavior of density fluctuations. In the case of a large magnetic field, if we look at equation \eqref{eq:Debye_dependence_of_electric_susceptibility} that shows the dependence of the electric susceptibility on the Debye length and the plasma fluctuation wavelength, we find that in the limit of small Debye length~(zero temperature limit), or in the limit where the wavelength of the plasma fluctuation is large, it approaches $H_{\pm}/\varepsilon_{\mathrm{L}}\rightarrow1$. For example, focusing on the electron density fluctuation equation \eqref{eq:density_fluctuation_of_ep}, in these limits, when considering the motion of a single electron, the other electron clouds make a response by shielding that motion, i.e., the non-collective term cancels out some of the collective terms. Eventually, the electron density fluctuations are dominated by the collective term that the positron cloud responds to when focusing on a single electron motion, and the net scattering effect turns out to be exactly the same as for a single particle.

\subsection{Why is the scattering not canceled out by the drift motion?}
When considering the scattering of electrons and positrons in a background magnetic field, one could think that the particles will drift in the same direction as the incident electromagnetic wave vector, leading to a cancellation of currents, and thus, scattering is almost negligible (see Appendix~\ref{Ap:pair_scattering}). However, when taking into account the scattering from density fluctuations in $e^{\pm}$ plasma (see Section~\ref{sec:Thomson_scattering_from_a_magnetized_plasma}), it becomes evident that the cross-section per particle remains unchanged compared to the case of considering single-particle scattering. Physically, the latter description is more accurate.

The difference in scattering intensities between the two pictures (Appendix~\ref{Ap:pair_scattering} vs. Section~\ref{sec:Thomson_scattering_from_a_magnetized_plasma}) arises from the distinct initial conditions of particles just before scattering at $t=0$. For instance, we focus on the differential cross-section attributed to the drift motion of particles in cold plasma. In the first scenario (Appendix~\ref{Ap:pair_scattering}), it is assumed that an electron and a positron are at rest and separated by a distance $d$ prior to scattering. The differential cross-section is estimated from equation \eqref{eq:current_pair} as
\begin{equation}
    \begin{aligned}
        \frac{\mathrm{d} \sigma^{(2)}_{\text{drift}}}{\mathrm{d} \Omega}&\propto\left|\widetilde{\boldsymbol{E}_{\text {rad }}}\right|^2\\
        &\propto(e^{i \boldsymbol{k} \cdot \boldsymbol{r}_{+}}-e^{i \boldsymbol{k} \cdot \boldsymbol{r}_{-}})(e^{-i \boldsymbol{k} \cdot \boldsymbol{r}_{+}}-e^{-i \boldsymbol{k} \cdot \boldsymbol{r}_{-}})\\
        &\propto2\left[1-\cos(k_xd)\right]\frac{\mathrm{d} \sigma^{(1)}_{\text{drift}}}{\mathrm{d} \Omega}.
    \end{aligned}
\end{equation}
The term involving different particles, i.e., the cross-term $e^{i \boldsymbol{k} \cdot (\boldsymbol{r}_{+}-\boldsymbol{r}_{-})}+\text{c.c.}$, takes a finite value and significantly alters the scattering intensity. On the other hand, when considering scattering from density fluctuations (Section~\ref{sec:Thomson_scattering_from_a_magnetized_plasma}), the squared absolute value of the radiative electric field is represented as in Equation \eqref{eq:total_electric_filed_abs}, and the differential cross-section is estimated as
\begin{equation}
    \begin{aligned}
        &\frac{\mathrm{d} \sigma^{(N_{+}+N_{-})}_{\text{drift}}}{\mathrm{d} \Omega}\propto\left\langle\left|\widetilde{\delta n_{+}}-\widetilde{\delta n_{-}}\right|^2\right\rangle_{\text{ensemble}}\\
        &\propto\left\langle\left\{\sum_{j=1}^{N_{-}} e^{i \boldsymbol{k} \cdot \boldsymbol{r}_j(t=0)} (\cdots)-\sum_{h=1}^{N_{+}} e^{i \boldsymbol{k} \cdot \boldsymbol{r}_h(t=0)} (\cdots)\right\}\right. \\
& \times\left.\left\{\sum_{s=1}^{N_{-}} e^{-i \boldsymbol{k} \cdot \boldsymbol{r}_s(t=0)} (\cdots)-\sum_{g=1}^{N_{+}} e^{-i \boldsymbol{k} \cdot \boldsymbol{r}_g(t=0)} (\cdots)\right\}\right\rangle_{\text {ensemble }}\\
&\propto  \left\langle\sum_{j=s}^{N_{+}}(\cdots)+\sum_{h=g}^{N_{-}}(\cdots)\right\rangle_{\text {ensemble }} \\
& -\left\langle\sum_{j \neq s}(\cdots)+\sum_{j, g}(\cdots)+\sum_{h, s}(\cdots)+\sum_{h \neq g}(\cdots)\right\rangle_{\text {ensemble }} \\
&\propto  \left(N_{+}+N_{-}\right) \frac{\dd \sigma^{(1)}_{\text{drift}}}{\dd \Omega}+0.
\end{aligned}
\end{equation}
Here, the cross-terms in the differential cross-section (the second row from the bottom) become zero when averaging over position and velocity, providing a correct depiction. The mistake in the first scenario is the failure to consider the statistical nature of plasma particles, which are distributed randomly following a particular distribution prior to scattering.

%
\subsection{Observational implications of scattering spectra}
\citet{2007ApJ...670..693H} showed that the interpulse has several spectral bands in their observations of the Crab pulsar. They also showed that the spacing between adjacent peak frequency bands of interpulse increases in proportion to frequency, i.e., $\nu\propto\Delta\nu$. The emission mechanism responsible for explaining such a spectrum is not fully understood.

CTS from plasma may explain the spectrum bands of the pulsar. As seen from Figure~\ref{fig_Collective_spectra}, when electromagnetic waves are scattered in magnetized thermal plasma, the scattered spectrum has peaks separated by the cyclotron frequency. Also, from the equation \eqref{eq:collective_total_cross_section_cold}, if the magnetic field is large enough, the total scattering cross-section depends on $\sigma\propto(\nu/\nu_{\text{c}})^2$. Hence, the higher-frequency electromagnetic waves are scattered more in the larger magnetic field region, resulting in larger peak separations at higher frequencies, which qualitatively agrees with their observation.

\subsection{Optical depth for fast radio bursts to induced Compton scattering}
Based on the discussion of Thomson scattering in magnetized plasma and the collective effect, we will evaluate the effective optical depth for induced Compton scattering of X-mode electromagnetic waves in $e^{\pm}$ plasma in a strong magnetic field.

It should be noted that the scattering cross-section derived from CTS is valid only within the range where plasma density fluctuations can be treated perturbatively. In situations with strong non-linear effects, such as induced Compton scattering of large amplitude electromagnetic waves, there is no guarantee that the differential scattering cross-section for CTS, given by equation \eqref{eq:Collective_cross_section}, can be applied. However, we will proceed with the discussion, assuming its applicability.

We simplify our analysis by providing an order-of-magnitude estimate for the effective optical depth of induced Compton scattering. Taking both the electron recoil and quantum corrections to the Thomson scattering cross-section into account, as indicated in previous studies~\citep{1976MNRAS.174...59B,1996AstL...22..399L}, we incorporate our calculations to the lowest order of Planck constant \(\hbar\). In order to avoid the substantial increase in complexity, our final expression omits to incorporate angular dependence.

In discussing induced Compton scattering, we first review the case without a magnetic field, then examine the magnetized case by applying the same framework. The effective cross-section for induced Compton scattering without a magnetic field is briefly derived. The Boltzmann equation for photons considering Compton scattering in the collision term is expressed as
\begin{equation}
\begin{aligned}
& \frac{\partial}{\partial t} N(\omega, \boldsymbol{\Omega})+c(\boldsymbol{\Omega} \cdot \nabla) N(\omega, \boldsymbol{\Omega}) \\
= & ~n_{\mathrm{e}} \int \mathrm{d}^3 \boldsymbol{k}^{\prime} \mathrm{d}^3 \boldsymbol{k}^{\prime \prime}\left[P\left(\boldsymbol{k}^{\prime} \rightarrow \boldsymbol{k}^{\prime \prime}\right)-P\left(\boldsymbol{k}^{\prime \prime} \rightarrow \boldsymbol{k}^{\prime}\right)\right] \delta\left(\boldsymbol{k}-\boldsymbol{k}^{\prime \prime}\right).
\end{aligned}
\label{eq:evolution_equation_for_photon_basic}
\end{equation}
Let $N(\omega, \boldsymbol{\Omega})$ be the photon occupation number at angular frequency $\omega$ in the direction $\bm{\Omega}=(\theta,\varphi)$, and $P\left(\bm{k}^{\prime} \rightarrow \bm{k}^{\prime \prime}\right)$ be the probability that an electromagnetic wave with wavenumber $\bm{k}^{\prime}$ transitions to $\bm{k}^{\prime \prime}$. The transition probability can be formulated in terms of Compton scattering as
\begin{equation}
\begin{aligned}
&P\left(\boldsymbol{k} \rightarrow \boldsymbol{k}^{\prime}\right) \mathrm{d}^3 \boldsymbol{k} \mathrm{d}^3 \boldsymbol{k}^{\prime}\\
&=c \mathrm{~d} \sigma N(\omega, \boldsymbol{\Omega})\left[1+N\left(\omega^{\prime}, \boldsymbol{\Omega}^{\prime}\right)\right] \delta\left(\omega^{\prime}-\omega+\Delta \omega\right) \mathrm{d} \omega^{\prime} \mathrm{d}^3 \boldsymbol{k},    
\end{aligned}
\label{eq:transition_probability_Compton}
\end{equation}
where $c~ \mathrm{d} \sigma N(\omega, \boldsymbol{\Omega})$ represents a scattered photon number flux, that is $\mathrm{d}\sigma$ represents a differential cross-section, and $\Delta \omega=\omega-\omega^{\prime}$ is frequency change due to Compton scattering.

First, we consider the low-energy limit of unpolarized Compton scattering without a background magnetic field
\begin{equation}
  \Delta \omega=\frac{\hbar \omega \omega^{\prime}}{m_{\mathrm{e}} c^2}(1-\cos \theta),  
\end{equation}
\begin{equation}
\frac{\mathrm{d} \sigma}{\mathrm{d} \Omega}=\frac{1}{2} r_{\mathrm{e}}^2\left(\frac{\omega^{\prime}}{\omega}+\frac{\omega}{\omega^\prime}-1+\cos ^2 \theta\right)+\mathcal{O}\left(\left(\frac{\hbar \omega}{m_{\mathrm{e}}c^2}\right)^2\right),
\label{eq:Thomson_cross-section_quantum_correction}
\end{equation}
where the first-order low-energy expansion of the Klein-Nishina's formula is employed as the differential cross-section. Then the evolution equation for photons due to induced Compton scattering is derived as \citep{1975Ap&SS..36..303B,1982MNRAS.200..881W}
\begin{equation}
\frac{\mathrm{d} N}{\mathrm{~d} t} \simeq \frac{3 \sigma_{\mathrm{T}}}{8 \pi} n_{\mathrm{e}} N \int \mathrm{d} \Omega^{\prime}\left(1+\cos ^2 \theta\right) \frac{\hbar}{m_{\mathrm{e}} c}(1-\cos \theta) \frac{\partial}{\partial\omega}\left(\omega^2 N^{\prime}\right).
\label{eq:basic_equation_for_induced_Compton_scattering}
\end{equation}
If the radiation is collimated, as in the case far from the source, the small factor $(1-\cos\theta)$ reduces the scattering within the beam. On the other hand, when the scattering from the beam into the background radiation outside the beam dominates, the scattering enhances the initially weak background exponentially \citep{1996AstL...22..399L}.

We define the effective optical depth as the amplification factor of the scattered light and substitute $N=N_0e^{\tau_{\text{ind}}}$ as the formal solution to the basic equation \eqref{eq:basic_equation_for_induced_Compton_scattering}. Then, the order evaluation reveals the following
\begin{equation}
    \frac{\tau_{\text {ind }}}{\Delta t}\sim\frac{3 \sigma_{\mathrm{T}}}{8 \pi} n_{\mathrm{e}}\Delta\Omega\frac{\hbar}{m_{\mathrm{e}} c}\frac{1}{\omega}\omega^2 N^{\prime},
\end{equation}
where $\Delta t$ and $\Delta\Omega$ are the pulse width and opening angle of the incident electromagnetic wave, respectively. The spectral flux at the scattering point and the isotropic luminosity are expressed by
\begin{equation}
\begin{aligned}
 F_{\nu}&\sim2\times\frac{2\pi\hbar}{c^2}\Delta\Omega~\nu^3N^{\prime}, \\
L_{\gamma}&\sim 4\pi r^2\nu F_{\nu}.
\end{aligned}
\end{equation}
Here $r$ is the distance from the center to the scattering point, and $\omega=2\pi\nu$. From the above, the effective optical depth of induced Compton scattering is evaluated in the following
\begin{equation}
\tau_{\text {ind }} \sim n_{\mathrm{e}}\sigma_{\mathrm{T}}c \Delta t\frac{ 3\pi L_{\gamma}  }{4r^2 m_{\mathrm{e}} \omega^3}.
\end{equation}
\\
\onecolumngrid
\rule{\linewidth}{0.5pt}

Next, we consider induced Compton scattering of X-mode waves in a strong magnetic field. We refer to the discussion by \citet{2000ApJ...540..907G} for the differential cross-section and frequency shift of Compton scattering in a strong magnetic field. We impose the following assumptions in deriving the details.
\begin{itemize} 
\item A uniform magnetic field $\bm{B}_0=(B_0,0,0)$ exists in the $x$-axis direction.
\item Consider an electron or positron in its rest frame and assume that the wave number vector of the incident electromagnetic wave is oriented in the direction of the magnetic field, i.e., the $x$ direction. The direction of this wave number vector is realized by the Lorentz aberration when the particle is in ultra-relativistic motion in a laboratory system, as viewed in the rest frame of particles.
\item Assume that the electron or positron is in the lowest Landau level.
\end{itemize}
In the strong magnetic field, the frequency shift of photons due to Compton scattering and the differential cross-section for photons polarized perpendicular to the background magnetic field can be expressed by \citep{2000ApJ...540..907G}
\begin{equation}
\Delta \omega=\omega-\frac{2 \omega}{1+\frac{\hbar\omega}{m_{\text{e}}c}\left(1-\cos \theta_{\text{B}}\right)+\sqrt{\left\{1+\frac{\hbar\omega}{m_{\text{e}}c}\left(1-\cos \theta_{\text{B}}\right)\right\}^2-2 \frac{\hbar\omega}{m_{\text{e}}c} \sin ^2 \theta_{\text{B}}}},
\end{equation}
\begin{equation}
\frac{\dd \sigma_{\perp}}{\dd \Omega} \approx \frac{3 \sigma_{\text{T}}}{32 \pi} \frac{\omega \omega^{\prime 2}}{\left(2 \omega-\omega^{\prime}\right)}\left(1+\cos ^2 \theta_{\text{B}}\right)\left[\frac{1}{\left(\omega-\omega_{\text{c}}\right)^2}+\frac{1}{\left(\omega+\omega_{\text{c}}\right)^2}\right],
\label{eq:cross-section_ultra_rela}
\end{equation}
where, $\omega$, $\omega^{\prime}$, and $\theta_{\text{B}}$ represent the angular frequencies of the incident photon, scattered photon, and the angle between the background magnetic field and the direction of the scattered photon, respectively. The frequency shift due to Compton scattering can be expanded up to the first order in the quantum correction parameter $\frac{\hbar\omega}{m_{\text{e}}c}$ as
\begin{equation}
\Delta \omega=\frac{\hbar \omega^2}{m_{\text{e}}c}\left\{\frac{1}{2}\left(1+\cos ^2 \theta_{\text{B}}\right)-\cos \theta_{\text{B}}\right\}+ \mathcal{O}\left(\left(\frac{\hbar \omega}{m_{\mathrm{e}}c^2}\right)^2\right).  
\label{eq:quantum_correction_strong_magnetic_field}
\end{equation}
Depending on whether a magnetic field is present, there is a significant difference in the treatment of the differential cross-section $\dd\sigma$ in equations \eqref{eq:evolution_equation_for_photon_basic} and \eqref{eq:transition_probability_Compton}.  Without a magnetic field, the quantum correction from the Klein-Nishina's formula appears in the second order, as seen in equation \eqref{eq:Thomson_cross-section_quantum_correction}. However, with a magnetic field, the quantum correction to the differential cross-section given in equation \eqref{eq:cross-section_ultra_rela} emerges in the first order, the same as the frequency shift. Therefore, we will consider quantum corrections to the differential cross-section up to the first order in the following discussion.

Substituting the differential cross-section \eqref{eq:cross-section_ultra_rela} and the frequency shift \eqref{eq:quantum_correction_strong_magnetic_field} into equation \eqref{eq:transition_probability_Compton}, the evolution equation for photons due to induced Compton scattering is derived as
\begin{equation}
\begin{aligned}
 \frac{\dd N}{\dd t} \simeq &\frac{3 n_{\text{e}}}{16 \pi} \sigma_{\text{T}}\left[\left(\frac{\omega}{\omega-\omega_{\text{c}}}\right)^2+\left(\frac{\omega}{\omega+\omega_{\text{c}}}\right)^2\right] N\\
 &\times\int\dd \Omega^{\prime}\left(1+\cos ^2 \theta_{\text{B}}\right) \frac{\hbar}{m_{\mathrm{e}} c}\left\{\frac{1}{2}\left(1+\cos ^2 \theta_{\text{B}}\right)-\cos \theta_{\text{B}}\right\}\frac{1}{\omega}\frac{\partial}{\partial\omega}\left(\omega^3 N^{\prime}\right). 
 \label{eq:evolution_equation_for_induced_Compton_magnetic_field}
\end{aligned}
\end{equation}
Focusing on the \( \theta_{\text{B}} \) dependence in equation \eqref{eq:evolution_equation_for_induced_Compton_magnetic_field}, we find that scattering from beam to beam is suppressed. On the other hand, scattering from the beam to the background radiation is exponentially amplified, becoming the dominant scattering, akin to the case without a magnetic field. Therefore, by estimating the order of magnitude, the effective optical depth for photons due to induced Compton scattering can be expressed by
\begin{equation}
\tau_{\mathrm{ind}}^{\perp}  \sim\left(\frac{\omega}{\omega_{\mathrm{c}}}\right)^2 n_{\mathrm{e}}\sigma_{\mathrm{T}}c \Delta t\frac{3\pi L_{\gamma}  }{4r^2 m_{\mathrm{e}} \omega^3}.
\end{equation}
The effective cross-section of the induced Compton scattering of X-mode waves is roughly suppressed by $(\omega/\omega_{\text{c}})^2$ in a strong magnetic field.
%
%
\twocolumngrid
\rule{\linewidth}{0.5pt}
\section{SUMMARY}
In this study, we estimated the Thomson scattering cross-section for X-mode waves in $e^{\pm}$ plasma, considering the collective effect in the presence of a background magnetic field. The results showed that the order of magnitude of the cross-section per particle remains unchanged compared to the case of single-particle scattering. In a strong magnetic field, the motion of electrons and positrons is dominated by drift motion, and one could think that the currents of electrons and positrons cancel with each other, and the scattering is suppressed significantly. However, since the plasma particles and density fluctuations follow a thermal distribution before the scattering, the correlation between different particles during scattering becomes negligible, and the scattering cancellation effect is absent.

Furthermore, it was revealed that the collective effect in $e^{\pm}$ plasma is not entirely canceled out when a background magnetic field is present, contrary to what was previously demonstrated in studies without a background magnetic field \citep{1992PhFlB...4.2669S}. The effects of density fluctuations can be separated into contributions arising from the motion of particles in the direction of the incident electric field and the drift motion. The contribution from the motion in the direction of the incident electric field exhibits complete cancellation of the collective effect, similar to the case without a background magnetic field. On the other hand, in the contribution arising from the drift motion, the behavior of density fluctuations is independent of the charge sign. Thus, the collective effect is not canceled.

In a background magnetic field, the radiation energy response of $e^{\pm}$ plasma differs significantly between curvature radiation and Thomson scattering of X-mode electromagnetic waves. This distinction is rooted in the directional relationship between the background magnetic field and the plasma's response. While plasma particles can move unrestrictedly parallel to the background magnetic field, their movement is confined within the Larmor radius in the perpendicular direction. The plasma's response aligns with the background magnetic field's direction for curvature radiation due to the trajectory of radiating particles along the curved field, leading to radiation suppression. In contrast, during Thomson scattering of X-mode waves, the plasma's response direction is perpendicular to the background field, limiting its ability to counteract the scattered wave and resulting in less radiation suppression.

Plotting the differential cross-section for X-mode waves in $e^{\pm}$ plasma following the Maxwellian distribution revealed that the scattering wave spectrum exhibits peaks separated by the cyclotron frequency if the scattering wave propagates nearly perpendicular to the background magnetic field. If observed in pulsars and FRBs, such distinctive spectral features could provide valuable information about the scattering region's plasma and magnetic field strength.

Moreover, considering the effects of background magnetic fields and the collective effect, we investigated the induced Compton scattering of X-mode waves in a strong magnetic field and cold plasma. For the first time, we treat the induced Compton scattering of electromagnetic waves polarized perpendicular to a strong magnetic field from the first principles. As a result, the effective optical depth was found to be suppressed by the square of the cyclotron frequency, specifically $(\nu/\nu_{\text{c}})^2$, compared to without the magnetic field.

When considering FRBs propagating through the magnetosphere of a magnetar, it is found that FRBs propagating as X-mode waves in pair plasma have an expanded region from which they can escape the magnetosphere due to the effects of the background magnetic field. Additionally, by taking into account the relativistic effects of the scattering medium, it is indicated that the required Lorentz factor is smaller than in cases without a magnetic field.
\begin{acknowledgments}
We want to express our gratitude to Shuichi Matsukiyo, Ryo Yamazaki, Shuta Tanaka, Masanori Iwamoto, and Shoma F. Kamijima for their invaluable feedback and insightful discussions regarding the physics presented in this paper. We also appreciate the constructive comments from Koutarou Kyutoku, and Wataru Ishizaki during our regular group meetings. The discussions with Takahiro Tanaka, Hidetoshi Omiya, Takafumi Kakehi, and Masaki Nishiura were instrumental in bringing this research to fruition. This work was supported by JST SPRING, Grant Number JPMJSP2110,
and MEXT/JSPS KAKENHI Grant Numbers 23H05430, 23H04900, 23H01172, 22H00130, 20H00158, 20H01901, 20H01904.
\end{acknowledgments}

\appendix

\section{Thomson scattering in electron-positron plasma when $\omega\ll \omega_{\text{c}},\omega_{\text{p}}$}
\label{Ap:plasma_reaction}
\citet{2004ApJ...600..872G} estimated the curvature radiation from charged particles moving along an infinitely strong curved magnetic field, taking into account the response of the plasma. The results showed that the radiation is suppressed by a factor of roughly $(\omega/\omega_{\text{p}})^2$ compared to curvature radiation in vacuum. In this chapter, we estimate how the response of the plasma corrects the radiation intensity in Thomson scattering of X-mode electromagnetic waves in a strong magnetic field.

We impose the following assumptions in deriving the details.
\begin{itemize} 
\item The $e^{\pm}$ plasma is uniformly distributed with a number density $n_{\text{e}}$.
\item A uniform magnetic field $\bm{B}_0=(B_0,0,0)$ exists in the $x$-axis direction.
\item Assume an electron as a scattering particle.
\item The X-mode wave is incident on the magnetic field at a wave-number vector $\bm{k}_0$ and angular frequency $\omega_0$.
\item The motion of a particle in the wave field is approximated as non-relativistic.
\end{itemize}

In plasma, the wave equation for the electromagnetic potential is modified from the case in vacuum when the response of the plasma is taken into account. Liénard–Wiechert potential, which describes radiation from charged particles in a vacuum, is not applicable. Therefore, to estimate the radiation intensity from a charged particle, the work done by the radiative electric field on the emitting particle must be calculated directly.

The wave equation for the electromagnetic potential is derived by considering the response of the plasma. First, the wave equation of the Fourier-transformed electromagnetic field potential can be written as
\begin{equation}
\begin{array}{l}
\left(k^2-\frac{\omega^2}{c^2}\right) \widetilde{\phi}(\boldsymbol{k}, \omega)=4 \pi \widetilde{\rho}(\boldsymbol{k}, \omega), \\
\left(k^2-\frac{\omega^2}{c^2}\right) \widetilde{\boldsymbol{A}}(\boldsymbol{k}, \omega)=\frac{4 \pi}{c} \widetilde{\boldsymbol{j}}(\boldsymbol{k}, \omega).
\end{array}
\end{equation}
We can substitute the current density and charge density produced by the plasma in the source term. The plasma current can be written as
\begin{equation}
\widetilde{\boldsymbol{j}}_{\text {plasma }}=\boldsymbol{\sigma} \cdot \widetilde{\boldsymbol{E}}=i \frac{\omega}{c} \boldsymbol{\sigma} \cdot \widetilde{\boldsymbol{A}}-i \boldsymbol{\sigma} \cdot \boldsymbol{k} \widetilde{\phi}.
\end{equation}
Here, $\boldsymbol{\sigma}$ is the conductivity tensor of the cold $e^{\pm}$ plasma
\begin{equation}
\boldsymbol{\sigma} \equiv i \frac{\omega_{\mathrm{p}}^2}{4 \pi \omega}\left(\begin{array}{ccc}
1 & 0 & 0 \\
0 & \frac{1}{1-u} & 0 \\
0 & 0 & \frac{1}{1-u}
\end{array}\right),
\end{equation}
where we define the following
\begin{equation}
u \equiv\left(\frac{\omega_{\mathrm{c}}}{\omega}\right)^2, \quad s \equiv\left(\frac{\omega_{\mathrm{p}}}{\omega}\right)^2.
\end{equation}
The charge density produced by the plasma is obtained from the continuity equation
\begin{equation}
\frac{\partial \rho}{\partial t}+\nabla \cdot \boldsymbol{j}=0 \Rightarrow \widetilde{\rho}_{\text {plasma }}=\frac{\boldsymbol{k}}{\omega} \cdot \widetilde{\boldsymbol{j}}_{\text {plasma }}.
\end{equation}
By moving the plasma response term to the left-hand side and the source term to the right-hand side, the wave equations can be written as follows
\begin{equation}
\begin{aligned}
&\left(k^2-\frac{\omega^2}{c^2}\right) \widetilde{\phi}-4 \pi i\left\{\frac{\boldsymbol{k} \cdot(\boldsymbol{\sigma} \cdot \widetilde{\boldsymbol{A}})}{c}-\frac{\boldsymbol{k} \cdot(\boldsymbol{\sigma} \cdot \boldsymbol{k})}{\omega} \widetilde{\phi}\right\}\\
&=4 \pi \widetilde{\rho}_{\text {particle }}, \\
&\left(k^2-\frac{\omega^2}{c^2}\right) \widetilde{\boldsymbol{A}}-\frac{4 \pi i}{c}\left\{\frac{\omega}{c} \boldsymbol{\sigma} \cdot \widetilde{\boldsymbol{A}}-\boldsymbol{\sigma} \cdot \boldsymbol{k} \widetilde{\phi}\right\}=\frac{4 \pi}{c} \widetilde{\boldsymbol{j}}_{\mathrm{particle}}.
\end{aligned}
\end{equation}

The above equation is a four-simultaneous equation for four-potential $\widetilde{A^\alpha} \equiv\left(\widetilde{\phi}, \widetilde{A}_x, \widetilde{A}_y, \widetilde{A}_z\right)$ with four-current $\widetilde{j}_{\text {particle }}^\alpha \equiv\left(c\widetilde{\rho}, \widetilde{j}_x, \widetilde{j}_y, \widetilde{j }_z\right)$ as the source term, and can be solved algebraically. As a radiating particle, we assume an electron oscillating in the X-mode wave field and the background magnetic field. Then, from equation \eqref{eq:velocity_of_moving_particles}, the four-current produced by the radiating particle is described up to the order of $(\omega_0/\omega_{\text{c}})^1$ as follows
\begin{equation}
\begin{aligned}
\widetilde{j}_z(\boldsymbol{k}, \omega) &\simeq\frac{2 \pi e^2 E_0}{m_{\mathrm{e}} \omega_{\mathrm{c}}} \delta\left(\omega-\omega_0\right),\\
\widetilde{j}_{\text{particle}}^\alpha & =\left(\frac{c k_z}{\omega}, 0,0,1\right) \widetilde{j}_z.
\end{aligned}
\label{eq:source_current}
\end{equation}

From the above, by solving for the four-potential, we can obtain the electromagnetic field produced by the oscillating particle, taking the response of the plasma into account. We are interested in the limit where the angular frequency of the incident electromagnetic wave is sufficiently smaller than the cyclotron and plasma frequencies. That is, the following simultaneous limits can be considered
\begin{equation}
\frac{1}{s}=\left(\frac{\omega}{\omega_{\text{p}}}\right)^2\ll 1, \quad \frac{1}{u}=\left(\frac{\omega}{\omega_{\text{c}}}\right)^2\ll 1.
\nonumber
\end{equation}
Define $\alpha$ as follows
\begin{equation}
    \alpha\equiv\left(\frac{\omega_{\text{c}}}{\omega_{\text{p}}}\right)^2=\frac{u}{s},
\end{equation}
and Taylor expansion in $1/s$ while keeping $\alpha$ constant. In the lowest order expansion, the radiative electric field is estimated as follows
\begin{equation}
\widetilde{\bm{E}}=\left(\begin{array}{c}
0 \\
i \dfrac{k_z k_y \frac{\omega}{c} \alpha^2}{\left\{k^2 \alpha-\frac{\omega^2}{c^2}(1+\alpha)\right\}\left\{k_x^2 \alpha-\frac{\omega^2}{c^2}(1+\alpha)\right\}} \\
i \dfrac{\frac{\omega}{c} \alpha\left\{\left(k_x^2+k_z{ }^2\right) \alpha-\frac{\omega^2}{c^2}(1+\alpha)\right\}}{\left\{k^2 \alpha-\frac{\omega^2}{c^2}(1+\alpha)\right\}\left\{k_x^2 \alpha-\frac{\omega^2}{c^2}(1+\alpha)\right\}} 
\end{array}\right)\frac{4 \pi}{c} \widetilde{j_z}.
\label{eq:scattered_electric_field}
\end{equation}

The energy radiated per unit time by a particle in the X-mode electromagnetic waves is equal to the work done by the radiative electric field on the oscillating particle if the radiative energy loss of the particle is ignored. In other words, the time-averaged power of the radiation from the oscillating particle is
\begin{eqnarray}
&&\left\langle P\right\rangle_{\text{T}}=-\left\langle\int \dd^3 \bm{r} \operatorname{Re}\left[j_z^{\text {part }}(\bm{r}, t)\right] \operatorname{Re}\left[E_z(\bm{r}, t)\right]\right\rangle_{\text{T}} \nonumber \\
 &=&-\frac{1}{4(2 \pi)^8} \int \dd^3 \bm{r} \dd^3\bm{k}^3 \dd^3\bm{k}^{\prime} \dd \omega \dd \omega^{\prime}\left[\left\langle e^{i\left\{\left(\bm{k}^{\prime}-\bm{k}\right) \cdot \bm{r}-\left(\omega^{\prime}-\omega\right) t\right\}}\right\rangle_{\text{T}} \right. \nonumber\\
 &&\times\widetilde{j_z}\left(\bm{k}^{\prime}, \omega^{\prime}\right) \widetilde{E}_z^*(\bm{k}, \omega)\nonumber\\
 &+&\left.\left\langle e^{-i\left\{\left(\bm{k}^{\prime}-\bm{k}\right) \cdot \bm{r}-\left(\omega^{\prime}-\omega\right) t\right\}}\right\rangle_{\text{T}} \widetilde{j}_z^*\left(\bm{k}^{\prime}, \omega^{\prime}\right)\widetilde{E}_z(\bm{k}, \omega)\right].
\end{eqnarray}

The integral is calculated using the following procedure: first, apply the delta function for $\omega$, $\omega^{\prime}$, $x$, $k_x^{\prime}$, $k_y^{\prime}$, $z$, and $k_z^{\prime}$. Second, perform complex integration for $k_y$. Finally, carry out the remaining two-dimensional integrals for $k_x$ and $k_z$. A concrete calculation process is shown below. Substituting the radiative electric field \eqref{eq:scattered_electric_field} and source current \eqref{eq:source_current}, the integration can be written as follows
\begin{eqnarray}
\left\langle\frac{\dd^2 P}{\dd \mu \dd \nu}\right\rangle_{\text{T}}&=&-i \frac{e^4 E_0^2 \omega_0^3}{8 \pi^2 m_{\mathrm{e}}^2 c^4 \omega_{\mathrm{c}}^2} \frac{\left(\mu^2+\nu^2\right)-\frac{\alpha+1}{\alpha}}{\nu^2-\frac{\alpha+1}{\alpha}} \int \dd y \delta(y) \nonumber\\
&\times& \int_{-\infty}^{+\infty} \dd k_y\left[\frac{e^{i k_y y}}{k_y^2-\left(\frac{\omega_0+i \varepsilon}{c}\right)^2\left(\frac{\alpha+1}{\alpha}-\mu^2-\nu^2\right)}\right.\nonumber\\
&-&\left.\frac{e^{i k_y y}}{k_y^2-\left(\frac{\omega_0-i \varepsilon}{c}\right)^2\left(\frac{\alpha+1}{\alpha}-\mu^2-\nu^2\right)}\right].
\nonumber
\end{eqnarray}
Here
\begin{equation}
\mu \equiv \frac{k_z c}{\omega_0}, \quad \nu \equiv \frac{k_x c}{\omega_0},
\end{equation}
are dimensionless quantities of $k_z$ and $k_x$, respectively. Applying the residue theorem to the $k_y$ integral, a case separation arises depending on the magnitude of $\mu$ and $\nu$ as follows
\begin{equation}
\begin{aligned}
 \left\langle\frac{\dd^2 P}{\dd \mu \dd \nu }\right\rangle_{\text{T}}&=-i \frac{e^4 E_0^2 \omega_0^3}{8 \pi^2 m_{\mathrm{e}}^2 c^4 \omega_{\text{c}}^2} \frac{\left(\mu^2+\nu^2\right)-\frac{\alpha+1}{\alpha}}{\nu^2-\frac{\alpha+1}{\alpha}}\\
 &\times\left\{\begin{array}{c}
\frac{2 \pi i c}{\omega_0} \frac{1}{\sqrt{\frac{\alpha+1}{\alpha}-\mu^2-\nu^2}} \quad\left(\mu^2+\nu^2 \leq \frac{\alpha+1}{\alpha}\right) \\\\
0\quad\left(\mu^2+\nu^2 \geq \frac{\alpha+1}{\alpha}\right)
\end{array}\right. .  
\end{aligned}
\nonumber
\end{equation}
Finally, by performing the remaining $\mu$ and $\nu$ integrals, the time-averaged radiative energy per unit time can be calculated as,
\begin{eqnarray}
\langle P\rangle_{\text{T}}&=&\frac{e^4 E_0^2}{4 \pi m_{\mathrm{e}}^2 c^3}\left(\frac{\omega_0}{\omega_{\text{c}}}\right)^2 \int_{\mu^2+\nu^2 \leq \frac{\alpha+1}{\alpha}} \dd \mu \dd \nu \frac{\sqrt{\frac{\alpha+1}{\alpha}-\mu^2-\nu^2}}{\frac{\alpha+1}{\alpha}-\nu^2}\nonumber\\
&=&\frac{e^4 E_0^2}{4 m_{\mathrm{e}}^2 c^3}\left(\frac{\omega_0}{\omega_{\text{c}}}\right)^2 \sqrt{1+\left(\frac{\omega_{\text{p}}}{\omega_{\text{c}}}\right)^2}\nonumber.
\end{eqnarray}

Next, we evaluate the scattering cross-section of the radiation process under consideration. In $e^{\pm}$ plasma, the group velocity of X-mode waves deviates from the speed of light, according to the dispersion relation of X-mode waves
\begin{equation}
    \frac{c^2k^2}{\omega^2}=1-\frac{s}{1-u}\xrightarrow{1/s,1/u\ll1}1+\frac{1}{\alpha}.
\end{equation}
Therefore, we can evaluate the group velocity of X-mode waves as follows
\begin{equation}
v_{\mathrm{g}}^{\mathrm{X}}=\frac{\mathrm{d} \omega}{\mathrm{d} k} \xrightarrow{1/s,1/u\ll1} c \frac{1}{\sqrt{1+\frac{1}{\alpha}}}.
\end{equation}
Hence, Thomson scattering cross-section with the X-mode electromagnetic wave considering the plasma response is evaluated as follows
\begin{equation}
\begin{aligned}
\sigma&=\frac{8 \pi}{v_{\mathrm{g}}^{\mathrm{X}} E_0^2} \cdot \frac{e^4 E_0^2}{4 m_{\mathrm{e}}^2 c^3}\left(\frac{\omega_0}{\omega_{\mathrm{c}}}\right)^2 \sqrt{1+\left(\frac{\omega_{\mathrm{p}}}{\omega_{\mathrm{c}}}\right)^2}\\
&=\frac{3}{4} \sigma_{\mathrm{T}}\left(\frac{\omega_0}{\omega_{\mathrm{c}}}\right)^2\left\{1+\left(\frac{\omega_{\mathrm{p}}}{\omega_{\mathrm{c}}}\right)^2\right\} .   
\end{aligned}
\end{equation}
The effect of plasma response can be evaluated as
\begin{equation}
\frac{3}{4} \left\{1+\left(\frac{\omega_{\text{p}}}{\omega_{\text{c}}}\right)^2\right\}\sim1,    
\end{equation} 
in the case of $\omega_{\text{c}}>\omega_{\text{p}}\gg\omega_0$ such as in the magnetar magnetosphere. This value is nearly similar to the vacuum case.

The effect of the response of $e^{\pm}$ plasma in the background magnetic field on the radiation energy is quite different between curvature radiation and Thomson scattering of X-mode electromagnetic waves. This can be interpreted physically by looking at the relationship between the direction of the background magnetic field and the direction in which the plasma responds. In the presence of a background magnetic field, plasma particles are free to move in the direction parallel to the background magnetic field but only within the Larmor radius in the direction perpendicular to the background magnetic field. In the case of curvature radiation, the response of the plasma is in the direction of the background magnetic field because the radiating particles are moving along the curved magnetic field. Hence, the plasma responds so that the radiation is canceled, thereby suppressing the radiation. On the other hand, in the case of Thomson scattering of X-mode electromagnetic waves, the plasma's direction of response is perpendicular to the background magnetic field, so the plasma cannot freely respond to the scattered wave, and the radiation is not suppressed so much.

\section{Misleading idea on Thomson scattering in $e^{\pm}$ plasma with a strong magnetic field}
\label{Ap:pair_scattering}

This section presents a misleading idea about the scattering in $e^{\pm}$ plasma with a strong magnetic field. Since drift motion dominates the motion of charged particles under the incident electromagnetic field in a strong magnetic field, one could think that electrons and positrons cancel out the scattering everywhere. This incorrect picture appears where initially static free electrons and free positrons simultaneously scatter X-mode waves. Specifically, consider Thomson scattering in $e^{\pm}$ plasma with a strong magnetic field in the following situation.
\begin{itemize} 
\item Suppose initially that an electron with charge $-e$ at $(-\frac{d}{2},0,0)$ and a positron with charge $+e$ at $(\frac{d}{2},0,0)$ are at rest. That is, suppose that the electron and positron are separated from each other by a microscopic distance $d$.
\item A uniform magnetic field $\bm{B}_0=(B_0,0,0)$ exists in the $x$-axis direction.
\item The X-mode wave is incident on the magnetic field at a wave-number vector $\bm{k}_0$ and angular frequency $\omega_0$.
\item The motion of a particle in the wave field is approximated as non-relativistic.
\end{itemize}

As in Appendix \ref{Ap:plasma_reaction}, determine the current density and electric field produced by the radiating particles. The current can be obtained by adding the electron and positron contributions
\begin{equation}
\boldsymbol{j}_{\mathrm{particle}}(\boldsymbol{r}, t)=\left[e \boldsymbol{v}_{+} \delta\left(x-\frac{d}{2}\right)-e \boldsymbol{v}_{-} \delta\left(x+\frac{d}{2}\right)\right] \delta(y) \delta(z).
\label{eq:current_pair}
\end{equation}
Then, from equation \eqref{eq:velocity_of_moving_particles}, substituting the motion of the particles, the current density is obtained as follows
\begin{equation}
\begin{aligned}
\widetilde{j_y}(\boldsymbol{k}, \omega) & =i \frac{4 \pi e^2 E_0 \omega_0}{m_{\mathrm{e}}\left(\omega_0^2-\omega_{\mathrm{c}}^2\right)} \cos \left(\frac{k_x d}{2}\right) \delta\left(\omega-\omega_0\right)\\
&\simeq i \frac{4 \pi e^2 E_0 \omega_0}{m_{\mathrm{e}}\left(\omega_0^2-\omega_{\mathrm{c}}^2\right)} \delta\left(\omega-\omega_0\right) \\
\widetilde{j_z}(\boldsymbol{k}, \omega) & =-i \frac{4 \pi e^2 E_0 \omega_0}{m_{\mathrm{e}}\left(\omega_0^2-\omega_{\mathrm{c}}^2\right)} \sin \left(\frac{k_x d}{2}\right) \delta\left(\omega-\omega_0\right) \\
& \simeq -i \frac{4 \pi e^2 E_0 \omega_0}{m_{\mathrm{e}}\left(\omega_0^2-\omega_{\mathrm{c}}^2\right)} \frac{k_x d}{2} \delta\left(\omega-\omega_0\right).
\end{aligned}
\end{equation}
Here, we approximated that the wavelength of the incident electromagnetic wave is sufficiently large compared to the average distance between the electrons and positrons, that is
\begin{equation}
k_xd\sim \frac{d}{\lambda}\ll1.    
\end{equation}

The electromagnetic potential is calculated to find the radiated electric field. The electromagnetic potential can be obtained by substituting the source into the following wave equation for the electromagnetic potential
\begin{equation}
 \left(k^2-\frac{\omega^2}{c^2}\right)\widetilde{\bm{A}}(\boldsymbol{k}, \omega)=\frac{4 \pi}{c} \widetilde{\bm{j}}_{\mathrm{particle }}.
\nonumber
\end{equation}
The radiated electric field is obtained as follows
\begin{equation}
\begin{aligned}
\widetilde{E_y}(\boldsymbol{k}, \omega)= &\frac{\left(k_y^2-\frac{\omega^2}{c^2}\right) \omega_0+k_y k_z \frac{k_x d}{2} \omega_{\mathrm{c}}}{\frac{\omega}{c}\left\{k^2-\frac{(\omega+i \varepsilon)^2}{c^2}\right\}}\\
&\times\frac{16 \pi^2 e^2 E_0}{m_{\mathrm{e}} c\left(\omega_0^2-\omega_{\mathrm{c}}^2\right)}\delta\left(\omega-\omega_0\right),
\end{aligned}
\end{equation}
\begin{equation}
\begin{aligned}
 \widetilde{E_z}(\boldsymbol{k}, \omega)=& -\frac{\left(k_z^2-\frac{\omega^2}{c^2}\right) \frac{k_x d}{2} \omega_{\mathrm{c}}+k_y k_z \omega_0}{\frac{\omega}{c}\left\{k^2-\frac{(\omega+i \varepsilon)^2}{c^2}\right\}} \\
 &\times i \frac{16 \pi^2 e^2 E_0}{m_{\mathrm{e}} c\left(\omega_0^2-\omega_{\mathrm{c}}^2\right)}\delta\left(\omega-\omega_0\right).
\end{aligned}
\end{equation}

The time-averaged energy radiated by an electron-positron pair per unit time can be obtained by estimating the work done by the radiative electric field on the oscillating particle as in Appendix \ref{Ap:plasma_reaction}
\begin{equation}
\begin{aligned}
 P^{(2)} & = P_y^{(2)}+ P_z^{(2)} \\
& =\frac{4 e^4 E_0^2}{3 m_{\mathrm{e}}^2 c^3} \frac{\omega_0^4}{\left(\omega_0^2-\omega_{\mathrm{c}}^2\right)^2}+\frac{2 e^4 E_0^2}{15 m_{\mathrm{e}}^2 c^3} \frac{\omega_0^2 \omega_{\mathrm{c}}^2}{\left(\omega_0^2-\omega_{\mathrm{c}}^2\right)^2}\left(\frac{\omega_0 d}{c}\right)^2.
\end{aligned}
\end{equation}
Eventually, the scattering cross-section per particle is obtained as follows
\begin{equation}
\sigma^{(1)}\simeq 2\left[\sigma_{\mathrm{T}}\left(\frac{\omega_0}{\omega_{\mathrm{c}}}\right)^4+\sigma_{\mathrm{T}}\left(\frac{\omega_0}{\omega_{\mathrm{c}}}\right)^2 \frac{1}{10}\left(\frac{\omega_0 d}{c}\right)^2\right].
\label{eq:cross_section_for_pair_particles}
\end{equation}

We give a physical interpretation of the scattering cross-section for an electron-positron pair scattering with an X-mode electromagnetic wave in a background magnetic field. The first term in the equation \eqref{eq:cross_section_for_pair_particles} represents the contribution of the incident electromagnetic wave oscillating in the electric field direction, and the second term represents the contribution of the particles oscillating in the drift direction. Since the motion in the drift direction does not depend on the charge sign, electrons and positrons cancel each other's current, and a factor $((\omega_0d)/c)^2$ appears in the scattering cross-section.

We estimate the scattering cancellation effect due to drift for FRBs of typical frequencies propagating through the magnetar magnetosphere. Assuming that the average interparticle distance between electrons and positrons is $d\sim n_{\text{e}}^{-1/3}\sim2\times10^{-5} ~\text{cm}$, the cyclotron frequency is $\omega_{\mathrm{c}} \sim 10^6~ \mathrm{GHz}$ and the FRB angular frequency is 1~GHz, the scattering suppression effect is estimated as follows
\begin{equation}
\sigma^{(1)}\sim\sigma_{\mathrm{T}}\left(\frac{\omega_0}{\omega_{\mathrm{c}}}\right)^2 \max \left\{10^{-12} \frac{\omega_9^2}{\omega_{\mathrm{c}, 15}^2}, 10^{-13} \frac{\omega_9^2}{d_{-5}^2}\right\}.   
\end{equation}

Thomson scattering in $e^{\pm}$ plasma appears significantly suppressed when the scattering cancellation effect due to drift motion is considered. However, as shown in this paper, when properly accounting for the particle statistics of the plasma, such cancellation effects do not occur~(see equation \eqref{eq:collective_total_cross_section_cold}).

\onecolumngrid
\newpage
\section{Detailed derivation of the positron/electron density fluctuation}
\label{Ap:derivation_of_density_fluctuation}
In this section, positron/electron density fluctuations in $e^{\pm}$ plasma are derived from the Vlasov equation.
\subsection{Vlasov equations for charged particles}
First, the Vlasov equation for the positron/electron distribution function is written down by separating the equilibrium state distribution function $F_{0\pm}$ and the first-order perturbation of the distribution function $\delta F_\pm$. The Vlasov equation for the positron/electron distribution function can be written as follows
\begin{equation}
   \frac{\partial F_{0\pm}}{\partial t}+\frac{\partial \delta F_\pm}{\partial t}+\bm{v} \cdot\left(\frac{\partial F_{0\pm}}{\partial \bm{r}}+\frac{\partial \delta F_\pm}{\partial \bm{r}}\right)\pm \frac{e}{m_{\text{e}}}\left(\bm{E}+\frac{\bm{v} \times \bm{B}_0}{c}\right) \cdot\left(\frac{\partial F_{0\pm}}{\partial \bm{v}}+\frac{\partial \delta F_\pm}{\partial \bm{v}}\right)=0.
\end{equation}
Considering the electric field $\bm{E}$ produced by the plasma and the perturbation $\delta F_\pm$ of the distribution function as perturbative quantities, the Vlasov equation can be divided into non-perturbative and perturbative components as follows
\begin{equation}
\begin{aligned}
&\frac{\partial F_{0\pm}}{\partial t}+\bm{v} \cdot \frac{\partial F_{0\pm}}{\partial \bm{r}} \pm \frac{e}{m_{\text{e}} c}(\bm{v} \times \bm{B}_0 )\cdot \frac{\partial F_{0\pm}}{\partial \bm{v}}=0,\\
\frac{\partial \delta F_{ \pm}}{\partial t}&+\bm{v} \cdot \frac{\partial \delta F_{ \pm}}{\partial \bm{r}} \pm \frac{e}{m_{\text{e}} c}\left(\bm{v} \times \bm{B}_0\right) \cdot \frac{\partial \delta F_{ \pm}}{\partial \bm{v}}\pm \frac{e}{m_{\text{e}}} \bm{E} \cdot \frac{\partial F_{0\pm}}{\partial \bm{v}}=0.
\end{aligned}
\label{eq:Vlasov_eq_for_ep}
\end{equation}
The second of \eqref{eq:Vlasov_eq_for_ep} is the equation to be satisfied by the first-order perturbation of the distribution function. The Lorentz force term, in the second equation \eqref{eq:Vlasov_eq_for_ep} can be simplified as follows by converting the differential variable from $\bm{v}$ to the angle $\varphi(t)$ between the velocity component perpendicular to the background magnetic field $\bm{B}_0=(B_0,0,0)$ and the $y$ axis~(see equation \eqref{eq:initial_condition_for_plasma})
\begin{equation}
\begin{aligned}
 \left(\bm{v} \times \bm{B}_0\right)\cdot \frac{\partial \delta F_{ \pm}}{\partial \bm{v}}=& v_z B_0 \frac{\partial \delta F_\pm}{\partial v_y}-v_y B_0 \frac{\partial \delta F_{ \pm}}{\partial v_z} \\
= & B_0\left\{v_z\left(\frac{v_y}{v_{ \perp}} \frac{\partial \delta F_{ \pm}}{\partial v_{\perp}}-\frac{v_z \cos ^2 \varphi}{v_y^2} \frac{\partial \delta F_\pm}{\partial \varphi}\right)-v_y\left(\frac{v_z}{v_{\perp}} \frac{\partial \delta F_{ \pm}}{\partial v_{\perp}}+\frac{\cos ^2 \varphi}{v_y} \frac{\partial \delta F_\pm}{\partial \varphi}\right)\right\} \\
= & -B_0 \frac{\partial \delta F_{ \pm}}{\partial \varphi}.
\end{aligned}
\end{equation}
In the end, the equation for the fluctuations of the distribution function is as follows
\begin{equation}
\frac{\partial \delta F_{ \pm}}{\partial t}+\bm{v} \cdot \frac{\partial \delta F_{ \pm}}{\partial \bm{r}} \mp \frac{e B_0}{m_{\text{e}} c} \frac{\partial \delta F_{ \pm}}{\partial \varphi} \pm \frac{e}{m_{\text{e}}} \bm{E} \cdot \frac{\partial F_{0\pm}}{\partial \bm{v}}=0.
\label{eq:Vlasov_eq_for_perturbation}
\end{equation}

\subsection{Derivation of the density fluctuations}
The equation \eqref{eq:Vlasov_eq_for_perturbation} for the fluctuations of the distribution function is Fourier-Laplace transformed for space and time as follows
\begin{equation}
\widetilde{\delta F}_{ \pm}(\bm{k}, \bm{v}, \omega)=\int_0^{\infty} \dd t ~e^{-i(\omega-i \varepsilon) t} \int \dd^3 \bm{r}~\delta F_{ \pm}(\bm{r}, \bm{v}, t) e^{i \bm{k} \cdot \bm{r}}.
\end{equation}
Then, the density fluctuations appearing in the spectral density function are represented by fluctuations in the distribution function as follows
\begin{equation}
\begin{aligned}
 \widetilde{\delta n_{ \pm}}(\bm{k}, \omega)&=\int_0^{\infty} \dd t ~e^{-i(\omega-i \varepsilon) t} \int \dd^3 \bm{r} ~\delta n_{ \pm}(\bm{r}, t) e^{i \boldsymbol{k} \cdot \bm{r}} \\
& =\int \dd^3 \bm{v} ~\widetilde{\delta F}_{ \pm}(\bm{k}, \bm{v}, \omega).
\end{aligned}
\end{equation}
Fourier-Laplace transform of equation \eqref{eq:Vlasov_eq_for_perturbation} leads to the following first-order differential equation for $\widetilde{\delta F_{ \pm}}$ with $\varphi$ as a variable
\begin{equation}
i(\omega-i \varepsilon-\bm{k} \cdot \bm{v}) \widetilde{\delta F_{ \pm}}(\bm{k}, \bm{v}, \omega) \mp \omega_{\text{c}} \frac{\partial \widetilde{\delta F_{ \pm}}(\bm{k}, \bm{v}, \omega)}{\partial \varphi}=\widetilde{\delta F_{ \pm}}(\bm{k}, \bm{v}, t=0) \mp \frac{e}{m_{\text{e}}} \widetilde{\bm{E}}(\bm{k}, \omega) \cdot \frac{\partial F_{0\pm}}{\partial \bm{v}}.
\end{equation}
This differential equation can be solved using the variation of constants
\begin{equation}
\begin{aligned}
& \widetilde{\delta F_{ \pm}}(\bm{k}, \bm{v}, \omega)=\frac{1}{\omega_{\text{c}}} \int \dd \varphi^{\prime} \exp \left[\mp\left(i \frac{\omega-i \varepsilon-k_x v_{\|}}{\omega_{\text{c}}} \varphi^{\prime}-i k_{\perp} r_{\text{L}} \sin \varphi^{\prime}\right)\right] \\
& \times\left\{ \mp \widetilde{\delta F_{ \pm}}(\bm{k}, \bm{v}, t=0)+\frac{e}{m_{\text{e}}} \widetilde{\bm{E}}(\bm{k}, \omega) \cdot \frac{\partial F_{0\pm}}{\partial \bm{v}}\right\} \exp \left[ \pm\left(i \frac{\omega-i \varepsilon-k_x v_{\|}}{\omega_{\text{c}}} \varphi-i k_{\perp} r_{\text{L}} \sin \varphi\right)\right].
\end{aligned}
\end{equation}

In the following, we will perform $\varphi^{\prime}$ integrals and represent the fluctuations of the distribution function using special functions. First, the exponential function is expanded using the infinite sum formula of the Bessel function
\begin{equation}
e^{i z \sin \varphi}=\sum_{l=-\infty}^{+\infty} J_l(z) e^{i l \varphi},
\label{eq:bessel_mugenwa_formula1}
\end{equation}
as follows
\begin{equation}
\begin{aligned}
&\widetilde{\delta F_{ \pm}}(\bm{k}, \bm{v}, \omega)=\frac{1}{\omega_{\text{c}}} \int \dd \varphi^{\prime} \exp \left[\mp\left(i \frac{\omega-i \varepsilon-k_x v_{\|}}{\omega_{\text{c}}} \varphi^{\prime}\right)\right] \sum_{l=-\infty}^{+\infty} J_l\left( \pm k_{\perp}r_{\text{L}}\right)e^{i l \varphi^{\prime}}\\  
& \times \left\{ \mp \widetilde{\delta F_{ \pm}}(\bm{k}, \bm{v}, t=0)+\frac{e}{m_{\text{e}}} \widetilde{\bm{E}}(\bm{k}, \omega) \cdot \frac{\partial F_{0\pm}}{\partial \bm{v}}\right\}\exp \left[ \pm\left(i \frac{\omega-i \varepsilon-k_x v_{\|}}{\omega_{\text{c}}} \varphi-i k_{\perp} r_{\text{L}} \sin \varphi\right)\right].
\end{aligned}
\label{eq:distribution_fluctuation_first}
\end{equation}
With respect to the second term in \eqref{eq:distribution_fluctuation_first}, we decompose the velocity derivative into directions perpendicular and parallel to the background magnetic field
\begin{equation}
\begin{aligned}
& \sum_{l=-\infty}^{+\infty} \int \dd \varphi^{\prime} \exp \left[\mp\left(i \frac{\omega-i \varepsilon-k_x v_{\|} \mp l\omega_{\text{c}}}{\omega_{\text{c}}} \varphi^{\prime}\right)\right]\widetilde{\bm{E}}(\bm{k}, \omega) \cdot \frac{\partial F_{0\pm}}{\partial \bm{v}} J_l\left( \pm k_{\perp}r_{\text{L}}\right) \\
& =\sum_{l=-\infty}^{+\infty}  \int \dd \varphi^{\prime} \exp \left[\mp\left(i \frac{\omega-i \varepsilon-k_x v_{\|} \mp l \omega_{\text{c}}}{\omega_{\text{c}}} \varphi^{\prime}\right)\right]\left(\widetilde{E_{\|}} \frac{\partial F_{0\pm}}{\partial v_{\|}}+\widetilde{E_{\perp}} \cos \varphi \frac{\partial F_{0\pm}}{\partial v_{\perp}}\right) J_l\left( \pm k_{\perp}r_{\text{L}}\right) \\
& = \pm i \sum_{l=-\infty}^{+\infty} \frac{\omega_{\text{c}}}{\omega-i \varepsilon-k_x v_{\|} \mp l \omega_{\text{c}}} \widetilde{E}_{\|} \frac{\partial F_{0\pm}}{\partial v_{\|}} J_l\left( \pm k_{\perp}r_{\text{L}}\right)\exp \left[\mp\left(i \frac{\omega-i \varepsilon-k_x v_{\|} \mp l \omega_{\text{c}}}{\omega_{\text{c}}} \varphi\right)\right] \\
& +\sum_{l=-\infty}^{+\infty} \int \dd \varphi^{\prime} \exp \left[\mp\left(i \frac{\omega-i \varepsilon-k_x v_{\|}}{\omega_{\text{c}}} \varphi^{\prime}\right)\right]\frac{J_{l+1}\left( \pm k_{\perp}r_{\text{L}}\right)+J_{l-1}\left( \pm k_{\perp}r_{\text{L}}\right)}{2} \widetilde{E}_{\perp} \frac{\partial F_{0\pm}}{\partial v_{\perp}} e^{i l \varphi^{\prime}} .
\end{aligned}
\end{equation}
Here, in the transformation from line 2 to line 3, the trigonometric functions $\cos\varphi$ were decomposed into exponential functions and absorbed into Bessel functions using equation \eqref{eq:bessel_mugenwa_formula1}. We now define the following differential operator
\begin{equation}
\widetilde{E}_{\|} \frac{\partial F_{0\pm}}{\partial v_{\| }} \pm \widetilde{E}_{\perp} \frac{\partial F_{0\pm}}{\partial v_{\perp}} \frac{l}{k_{\perp} r_{\text{L}}}\equiv\widetilde{\bm{E}} \cdot \frac{\partial F_{0\pm}}{\partial\bm{v}^*}.
\end{equation}
Using the Bessel function formula
\begin{equation}
J_{l+1}(z)+J_{l-1}(z)=\frac{2 l}{z} J_l(z),
\end{equation}
the fluctuations of the distribution function can be organized as follows
\begin{equation}
\widetilde{\delta F_{ \pm}}(\bm{k}, \bm{v}, \omega)=-i \sum_{l=-\infty}^{+\infty}  \sum_{m=-\infty}^{+\infty} \frac{J_l\left( \pm k_{\perp}r_{\text{L}}\right) J_m\left( \pm k_{\perp}r_{\text{L}}\right) e^{i(l-m) \varphi}}{\omega-i \varepsilon-k_x v_{\|} \mp l \omega_{\text{c}}} \left\{\widetilde{\delta F_{ \pm}}(\bm{k}, \bm{v}, t=0) \mp \frac{e}{m_{\text{e}}} \widetilde{\bm{E}} \cdot \frac{\partial F_{0\pm}}{\partial\bm{v}^*}\right\}.
\end{equation}
Hence, the positron/electron density fluctuation can be written as
\begin{equation}
\widetilde{\delta n_{ \pm}}(\bm{k}, \omega)=-i \sum_{l,m}\int \mathrm{d}^3 \boldsymbol{v}\frac{J_l\left( \pm k_{\perp}r_{\text{L}}\right) J_m\left( \pm k_{\perp}r_{\text{L}}\right) e^{i(l-m) \varphi}}{\omega-i \varepsilon-k_x v_{\|} \mp l \omega_{\text{c}}}\left\{\widetilde{\delta F_{ \pm}}(\bm{k}, \bm{v}, t=0) \mp \frac{e}{m_{\text{e}}} \widetilde{\bm{E}} \cdot \frac{\partial F_{0\pm}}{\partial\bm{v}^*}\right\}.
\label{eq:distribution_fluctuation_second}
\end{equation}

Then, for the first term of \eqref{eq:distribution_fluctuation_second}, the fluctuations of the distribution function just before the plasma scatters the electromagnetic waves~(at time $t=0$) can be written in terms of individual particle-like representations as follows
\begin{equation}
\begin{aligned}
\widetilde{\delta F_{ \pm}}(\bm{k}, \bm{v}, t=0) & =\sum_{j=1}^{N_{ \pm}} \int \dd^3 \bm{r} ~e^{i \bm{k} \cdot \bm{r}} \delta\left(\bm{r}-\bm{r}_{ \pm j}(0)\right) \delta(\bm{v}-\bm{v}_{\pm j}(0)) \\
& =\sum_{j=1}^{N_{ \pm}} e^{i \bm{k} \cdot \bm{r}_{ \pm j}(0)} \delta(\bm{v}-\bm{v}_{\pm j}(0)).
\end{aligned}
\end{equation}
Substituting this into the expression \eqref{eq:distribution_fluctuation_second}, we can write
\begin{equation}
\begin{aligned}
\widetilde{\delta n_{ \pm}}(\boldsymbol{k}, \omega) 
&=-i \sum_{j=1}^{N_{ \pm}} \sum_{l,m} e^{i \bm{k} \cdot \bm{r}_{ \pm}(0)} \frac{J_l( \pm k_{\perp} r_{\text{L}}) J_m\left(\pm k_{\perp} r_{\text{L}}\right) e^{i(l-m) \varphi_j(0)}}{\omega-i \varepsilon-k_x v_{\|} \mp l \omega_{\text{c}}}\\
&\pm \frac{4 \pi e}{m_{\text{e}} k^2} \sum_{l,m} \int \mathrm{d}^3 \bm{v} \frac{J_l\left( \pm k_{\perp} r_{\mathrm{L}}\right) J_m\left( \pm k_{\perp} r_{\mathrm{L}}\right) e^{i(l-m) \varphi}}{\omega-i \varepsilon-k_x v_{\|} \mp l \omega_{\mathrm{c}}}\widetilde{\rho}(\bm{k}, \omega) \bm{k} \cdot \frac{\partial F_{0\pm}}{\partial \bm{v}^*}.
\end{aligned}
\label{eq:distribution_fluctuation_third}
\end{equation}
Note that the Maxwell equation $\widetilde{\bm{E}}=\frac{4\pi i}{k^2}\bm{k}\widetilde{\rho}$ is applied to the second term in \eqref{eq:distribution_fluctuation_second}. In the velocity integral in the second term of \eqref{eq:distribution_fluctuation_third}, the following orthogonal relation from the angle integral
\begin{equation}
\int_0^{2 \pi} e^{i(l-m) \varphi}\dd\varphi=\left\{\begin{array}{cc}
2 \pi & l=m \\
0 & l \neq m
\end{array}\right.
\label{eq:orthogonal_relation_exponential}
\end{equation}
can be used to simplify the infinite sum of Bessel functions for $l$ and $m$. Using the equation for the electric susceptibility of electrons/positrons in magnetized plasma~(equivalent to equation \eqref{eq:electric_susceptibility_thermal}) 
\begin{equation}
H_{ \pm}(\bm{k}, \omega) \equiv\int \dd^3 \bm{v} \frac{4 \pi e^2 n_{\text{e}}}{m_{\text{e}} k^2} \sum_{l=-\infty}^{+\infty} \bm{k} \cdot \frac{\partial f_{ \pm}}{\partial \bm{v}^*} \frac{J_l^2\left( \pm k_{\perp} r_{\text{L}}\right)}{\omega-i \varepsilon-k_x v_{\|} \mp l \omega_{\text{c}}},
\label{eq:electric_susceptibility}
\end{equation}
the density fluctuations can be organized as follows
\begin{equation}
\begin{aligned}
\widetilde{\delta n_{ \pm}}(\boldsymbol{k}, \omega)
= -i \sum_{j=1}^{N_{ \pm}} \sum_{l,m} e^{i \boldsymbol{k} \cdot \boldsymbol{r}_{ \pm j}(0)} \frac{J_l\left( \pm k_{\perp} r_{\mathrm{L}}\right) J_m\left( \pm k_{\perp} r_{\mathrm{L}}\right) e^{i(l-m) \varphi_j(0)}}{\omega-i \varepsilon-k_x v_{\|} \mp l \omega_{\mathrm{c}}}\mp\frac{H_{ \pm}(\bm{k}, \omega)}{e} \widetilde{\rho}(\bm{k}, \omega).
\end{aligned}
\label{eq:distribution_fluctuation_forth}
\end{equation}
Finally, the charge density can be obtained in a self-consistent manner because the charge density is expressed as a linear combination of electron-positron density fluctuations as follows
\begin{equation}
\widetilde{\rho}(\bm{k}, \omega)=e \widetilde{\delta n_{ +}}(\bm{k}, \omega)-e \widetilde{\delta n_{ -}}(\bm{k}, \omega).
\end{equation}
The longitudinal dielectric tensor is defined by the electric susceptibility of the magnetized plasma as follows
\begin{equation}
    \varepsilon_{\mathrm{L}}(\boldsymbol{k}, \omega)=1+H_{+}(\boldsymbol{k}, \omega)+H_{ -}(\boldsymbol{k}, \omega).
\end{equation}
Then the charge density is expressed as follows
\begin{equation}
\begin{aligned}
\widetilde{\rho}(\boldsymbol{k}, \omega) 
&=-i\frac{e}{\varepsilon_{\mathrm{L}}}\left[i \sum_{j=1}^{N_{+}} \sum_{l,m} e^{i \boldsymbol{k} \cdot \bm{r}_{+j}(0)} \frac{J_l\left(  k_{\perp} r_{\mathrm{L}}\right) J_m\left( k_{\perp} r_{\mathrm{L}}\right) e^{i(l-m) \varphi_j(0)}}{\omega-i \varepsilon-k_x v_{\|} - l \omega_{\mathrm{c}}}\right.\\
&\left.- \sum_{h=1}^{N_{-}} \sum_{l,m}e^{i \boldsymbol{k} \cdot \bm{r}_{-h}(0)} \frac{J_l\left( -k_{\perp} r_{\mathrm{L}}\right) J_m\left( -k_{\perp} r_{\mathrm{L}}\right) e^{i(l-m) \varphi_h(0)}}{\omega-i \varepsilon-k_x v_{\|} + l \omega_{\mathrm{c}}}\right].
\end{aligned}
\end{equation}
Substituting this again into equation \eqref{eq:distribution_fluctuation_forth}, we obtain the final expression for the density fluctuation
\begin{equation}
\begin{aligned}
 \widetilde{\delta n_{\pm}}(\boldsymbol{k}, \omega) 
&=-i\left[\left(1-\frac{H_{\pm}}{\varepsilon_{\mathrm{L}}}\right) \sum_{j=1}^{N_{\pm}} e^{i \boldsymbol{k} \cdot \boldsymbol{r}_{\pm j}(0)} \sum_{l, m} \frac{J_l\left(\pm k_{\perp} r_{\text{L}}\right) J_m\left(\pm k_{\perp} r_{\text{L}}\right)}{\omega-i\varepsilon-k_x v_{\| }\mp l \omega_{\mathrm{c}}} e^{i(l-m) \phi_{0j}}\right. \\
& \left.+\frac{H_{\pm}}{\varepsilon_{\mathrm{L}}} \sum_{h=1}^{N_{\mp}} e^{i \boldsymbol{k} \cdot \boldsymbol{r}_{\mp h}(0)} \sum_{l, m} \frac{J_l\left(\mp k_{\perp} r_{\text{L}}\right) J_m\left(\mp k_{\perp} r_{\text{L}}\right)}{\omega-i\varepsilon-k_x v_{\| }\pm l \omega_{\mathrm{c}}} e^{i(l-m) \phi_{0h}}\right],
\end{aligned}
\end{equation}
where $\varphi(0)=\phi_0$ is the initial phase of particles~(see equation \eqref{eq:initial_condition_for_plasma}).
\twocolumngrid
\nocite{*}

\bibliography{apssamp}

\end{document}